\documentclass[preprint,12pt]{elsarticle}

\usepackage{amssymb}
\usepackage{graphicx}
\usepackage{subcaption}
\usepackage{caption}
\usepackage{float}
\usepackage{soul}
\usepackage{amsmath}
\usepackage{xcolor}
\usepackage{tabularx}
\usepackage{placeins}
\usepackage{tabularx,booktabs}
\usepackage{nomencl}
\usepackage{fancyhdr}
\usepackage[most]{tcolorbox}
\usepackage[hidelinks]{hyperref}

\journal{Int. J. Numer. Methods Heat Fluid Flow}

\newcommand{\AAMheader}{%
\begin{tcolorbox}[colback=black!3,colframe=black,title={Author Accepted Manuscript (AAM)}]
\textbf{Citation:} This is the author accepted manuscript for the article published as: \\
AmirHossein Ghaemi, Abbas Ebrahimi, Majid Hajipour, Seyyed Mohammad Mahdy Shobeiry, Arash Fath Lipaei, 
``Model Predictive and Reinforcement Learning Methods for Active Flow Control of an Airfoil with Dual-point Excitation of Plasma Actuators'', 
\textit{International Journal of Numerical Methods for Heat \& Fluid Flow} (Emerald), 2025.\\
\textbf{DOI:} \href{https://doi.org/10.1108/HFF-02-2025-0118}{10.1108/HFF-02-2025-0118}\\
\textbf{Accepted:} 26~June~2025\\
\textbf{Version:} Author Accepted Manuscript (AAM) -- not the final published version.\\
\textbf{Publisher’s version:} The final Version of Record is available at Emerald Insight: \href{https://www.emerald.com/insight/content/doi/10.1108/HFF-02-2025-0118}{link}.\\
\textbf{Abstract page:} \href{https://www.emerald.com/hff/article-abstract/doi/10.1108/HFF-02-2025-0118/1272439/Model-predictive-and-reinforcement-learning?redirectedFrom=fulltext}{Emerald Abstract Link}.\\
\textbf{Licence for deposit:} This AAM is made available under the Creative Commons Attribution–NonCommercial 4.0 International Licence (CC BY-NC 4.0): \href{https://creativecommons.org/licenses/by-nc/4.0/}{https://creativecommons.org/licenses/by-nc/4.0/}.
\end{tcolorbox}%
}

\makeatletter
\def\ps@pprintTitle{%
  \let\@oddhead\@empty
  \let\@evenhead\@empty
  \let\@oddfoot\@empty
  \let\@evenfoot\@empty
}
\makeatother

\begin{document}

\AAMheader

\begin{frontmatter}

\title{Model Predictive and Reinforcement Learning Methods for Active Flow Control of an Airfoil with Dual-point Excitation of Plasma Actuators}

\begin{abstract}

\noindent \textbf{Purpose} – This study aims to investigate the effectiveness of Model Predictive Control (MPC) and Reinforcement Learning (RL) approaches for active flow control over a NACA 4412 airfoil near the static stall condition at a Reynolds number of \(4 \times 10^{5}\). By systematically evaluating these control strategies, the research seeks to address a critical gap in optimizing excitation frequency and improving response time in flow control applications. The study contributes to a deeper understanding of the adaptability and performance of RL-based methods compared to traditional MPC in aerodynamic flow separation control.  

\noindent \textbf{Methodology} – The study employs a quantitative approach through numerical simulations of the Reynolds Averaged Navier-Stokes (RANS) equations with the Scale-Adaptive Simulation (SAS) turbulence model. Dielectric Barrier Discharge (DBD) plasma actuators, operating in dual-point excitation mode, are utilized for flow separation control. The research evaluates adaptive MPC, temporal difference reinforcement learning (TDRL), and deep Q-learning (DQL) in optimizing excitation frequency and expediting the stabilization process. Additionally, an integrated approach combining signal processing with DQL is examined to enhance control performance.  

\noindent \textbf{Findings} – This study explores advanced control strategies for optimizing aerodynamic performance by managing flow separation using plasma actuators. We evaluate adaptive MPC, TDRL, DQL, and DQL with signal processing, utilizing dual-point excitation via DBD plasma actuators. \textcolor{black}{Adaptive MPC successfully achieved a target lift coefficient (\(C_l\)) of 1.60 using an excitation frequency of approximately 110\,Hz, but struggled to reach higher target \(C_l\) values near the physical limits.} RL methods effectively optimized excitation frequencies, achieving a lift coefficient of approximately 1.62 in under 2.5 seconds with an excitation frequency of 100 or 200 Hz.

\noindent \textbf{Originality} – This study presents a novel comparison of RL and MPC methods for active flow control, utilizing DBD plasma actuators to mitigate flow separation and enhance aerodynamic performance. Prior approaches have primarily focused on either MPC or RL independently, often relying on offline learning with separate training and testing phases. In contrast, our research employs an online learning framework, where RL-based techniques such as TDRL, DQL, and signal processing-enhanced DQL dynamically adapt to real-time aerodynamic conditions. By simultaneously evaluating adaptive MPC and RL methods in an online learning setup, we provide new insights into their comparative performance in dynamic environments.

\end{abstract}

\begin{keyword}
Closed-loop active flow control \sep DBD plasma actuators \sep Flow separation control \sep Model predictive control \sep Reinforcement learning
\end{keyword}

\end{frontmatter}

\makenomenclature

\renewcommand{\nomname}{\textcolor{black}{Nomenclature}}
\renewcommand{\nomlabel}[1]{\textcolor{black}{#1}\hfil}
\renewcommand{\nomentryend}{\hspace*{0pt}}

\printnomenclature

\nomenclature{$c$}{Airfoil chord length [m]}
\nomenclature{$C_d$}{Drag coefficient [—]}
\nomenclature{$C_l$}{Lift coefficient [—]}
\nomenclature{$C_p$}{Pressure coefficient [—]}
\nomenclature{$E$}{Electric‑field magnitude [V·m$^{-1}$]}
\nomenclature{$e(t)$}{Zero‑mean white noise [—]}
\nomenclature{$E_0$}{Maximum electric field [V·m$^{-1}$]}
\nomenclature{$f$}{Actuator excitation frequency [Hz]}
\nomenclature{$F^{+}$}{Normalized frequency $f c / U_\infty$ [—]}
\nomenclature{$G$, $F$}{Convolution matrices [—]}
\nomenclature{$g_k$}{Markov parameters [—]}
\nomenclature{$\hat{y}(t+j|t)$}{$j$‑step prediction [—]}
\nomenclature{$J$}{MPC cost function [—]}
\nomenclature{$k(t)$}{RLS gain [—]}
\nomenclature{$k_1, k_2$}{Field‑gradient constants [—]}
\nomenclature{$n_a$}{Order of denominator polynomial [—]}
\nomenclature{$n_b$}{Order of numerator polynomial [—]}
\nomenclature{$N_p$}{Prediction horizon [—]}
\nomenclature{$N_u$}{Control horizon [—]}
\nomenclature{$P$}{Static pressure [Pa]}
\nomenclature{$q(s,a)$}{Action–value [—]}
\nomenclature{$Re$}{Reynolds number [—]}
\nomenclature{$t$}{Time [s]}
\nomenclature{$T_c$}{Controller time step [s]}
\nomenclature{TimeStep}{CFD solver time step [s]}
\nomenclature{$u$}{Control input (duty cycle) [Hz]}
\nomenclature{$u_i$}{Velocity component [m·s$^{-1}$]}
\nomenclature{$U_\infty$}{Free‑stream velocity [m·s$^{-1}$]}
\nomenclature{$U_e$}{Edge velocity (Wake flow) [m·s$^{-1}$]}
\nomenclature{$y^{+}$}{Wall‑unit normal coordinate [—]}
\nomenclature{$x, y$}{Cartesian coordinates [m]}
\nomenclature{$y(t)$}{System output ($C_l$) [—]}
\nomenclature{$z$}{Z transform [—]}
\nomenclature{$\vec{F}$}{Plasma body‑force density [N·m$^{-3}$]}
\nomenclature{$\alpha$}{Learning rate (RL) [—]}
\nomenclature{$\Delta$}{Backward‑difference operator [—]}
\nomenclature{$\Delta T$}{Discharge duration [s]}
\nomenclature{$\epsilon$}{Exploration rate (RL) [—]}
\nomenclature{$\gamma$}{Discount factor (RL) [—]}
\nomenclature{$\kappa$}{Exploration‑decay factor [—]}
\nomenclature{$\lambda$}{Control weight [—]}
\nomenclature{$\mu$}{Dynamic viscosity [Pa·s]}
\nomenclature{$\rho$}{Fluid density [kg·m$^{-3}$]}
\nomenclature{$\sigma_{ij}$}{Molecular viscous stress [Pa]}
\nomenclature{$\tau_{ij}$}{Total stress tensor [Pa]}
\nomenclature{$\tau_{ij}^T$}{Reynolds stress tensor [Pa]}
\nomenclature{$\theta$}{Parameter vector (RLS) [—]}
\nomenclature{$\varphi(t)$}{Regressor vector (RLS) [—]}
\nomenclature{$\delta_{ij}$}{Kronecker delta [—]}
\nomenclature{$\vartheta$}{AC‑voltage frequency [Hz]}

\nomenclature{AFC}{Active Flow Control [—]}
\nomenclature{ARIMAX}{Auto‑regressive integrated moving average (exog.) [—]}
\nomenclature{CARIMA}{Controller‑ARIMAX model [—]}
\nomenclature{CFC}{Closed‑loop Flow Control [—]}
\nomenclature{CFD}{Computational Fluid Dynamics [—]}
\nomenclature{DBD}{Dielectric Barrier Discharge [—]}
\nomenclature{DQL}{Deep‑Q Learning [—]}
\nomenclature{LES}{Large‑Eddy Simulation [—]}
\nomenclature{LSTM}{Long Short‑Term Memory [—]}
\nomenclature{MPC}{Model Predictive Control [—]}
\nomenclature{RANS}{Reynolds‑Averaged Navier–Stokes [—]}
\nomenclature{RL}{Reinforcement Learning [—]}
\nomenclature{SAS}{Scale‑Adaptive Simulation [—]}
\nomenclature{S$_{SAS}$}{SAS‑specific source term [—]}
\nomenclature{TDRL}{Temporal‑Difference Reinforcement Learning [—]}


\FloatBarrier
\section{\label{sec:Intro}Introduction}

Closed-loop flow control (CFC) methods have emerged as a promising solution for mitigating adverse hydrodynamic and aerodynamic effects. These methods dynamically monitor and influence the flow field, based on feedback from real-time flow conditions, utilizing sensors and actuators \cite{liu2024primary, ding2024effects}. Implementing CFC methods enables more efficient and adaptable hydro/aerodynamic systems across various engineering fields. In aviation, CFC can enhance aircraft maneuverability, delay stall, and improve fuel efficiency \cite{A5}. The automotive industry also benefits from these techniques through drag reduction and improved vehicle stability \cite{eulalie2018active}. \textcolor{black}{ In wind energy applications, CFC contributes to increased turbine power output \cite{rezaeiha2019active}, mitigates wake effects to boost overall wind farm performance \cite{bay2019flow}, and helps reduce noise emissions from turbines \cite{ahmed2025innovative}. Additionally, CFC can alleviate structural loads \cite{macquart2017decoupling}, thereby reducing fatigue and extending the operational lifespan of wind turbines.}

\textcolor{black}{A critical area where CFC demonstrates significant value is in managing flow separation, a fundamentally nonlinear and high-dimensional phenomenon that leads to substantial performance losses, such as increased drag, reduced lift, and flow-induced vibrations \cite{das2013unsteady, hajipour2022active}. Flow separation primarily arises due to adverse pressure gradients that decelerate boundary layer flow, causing detachment and resulting in complex interactions among turbulence, boundary layer dynamics, and unsteady vortical structures \cite{simpson1989turbulent, eich2020large}. Traditional open-loop control strategies often fall short in addressing these complexities due to their inability to adapt to time-varying flow conditions and disturbances \cite{hajipour2022active}. In contrast, CFC leverages real-time sensor feedback, coupled with actuators and adaptive control algorithms, to dynamically influence flow characteristics, significantly mitigating or even eliminating separated flows \cite{kotapati2010nonlinear, deem2020adaptive}. As a result, CFC offers robust and efficient solutions that are essential for managing separated flows in engineering applications. Accordingly, recent research in CFC has increasingly focused on developing novel approaches to improve aerodynamic performance. The following section reviews some of the most relevant contributions in this domain.}

Closed-loop flow control methods can be categorized into two main groups: model-based and data-driven approaches. To date, various model-based methods have significantly advanced the development of flow control strategies \textcolor{black}{\cite{ransing2024guest, ghalambaz2024physics}. Rajesh et al. \cite{rajesh2025dynamic} investigated surface modifications on a circular cylinder with ridges of varying configurations, employing dynamic mode decomposition (DMD) for flow field analysis and a gated recurrent unit (GRU) deep learning model to predict surface pressure distributions efficiently}. Proctor et al. \cite{5} extended DMD by incorporating control effects, resulting in precise input-output models for complex, high-dimensional systems, referred to as dynamic mode decomposition with control (DMDc). Addressing uncertainties in system identification, Dovetta et al. \cite{13} proposed a hybrid framework that combines perturbative methods with Monte Carlo techniques to estimate the statistical properties of identified systems efficiently. Huang and Kim \cite{14} tackled separated flows using a linear optimal control strategy based on system identification, achieving a significant reduction in separation bubble size. Moreover, Obeid et al. \cite{15} applied the NARMAX identification technique and a Proportional–Integral (PI) controller for closed-loop control of flow separation over a NACA 0015 airfoil, markedly improving the lift coefficient and demonstrating the effectiveness of advanced system identification methods.

Data-driven approaches and deep reinforcement learning (DRL) have recently attracted much attention in the study of flow separation control \cite{ishize2024flow, hachem2024reinforcement}. Vignon et al. \cite{9} provided an overview of DRL frameworks in active flow control, identifying limitations and future directions. Shimomura et al. \cite{2} showed the effectiveness of a CFC system using a deep Q Network (DQN) on a NACA 0015 airfoil, achieving higher control gains accompanied by the attached flow. A RL-based method for actuator selection was introduced by Paris et al. \cite{3} to optimize control performance of laminar flow separation around an airfoil. Additionally, Vinuesa et al. \cite{7} and Varela et al. \cite{8} reviewed DRL applications in flow control, with the aim of highlighting the potential for drag reduction of wings and circular cylinders. DRL has been successfully applied to active flow control (AFC) in turbulent conditions, achieving a 30\% drag reduction at a Reynolds (Re) number of 1000 by reducing turbulent fluctuations and elongating the recirculation bubble \cite{17}. In varying Re number flows around a square cylinder, DRL's adaptability was demonstrated with drag reductions up to 47\%, effectively suppressing vortex shedding across different Re values \cite{18}. In the control of turbulent separation bubbles (TSB), DRL outperformed classical periodic forcing, achieving a 25.3\% reduction in TSB length at a friction Reynolds number of 180 \cite{19}. At a friction Reynolds number of 750, DRL continued to show superiority with an 8.9\% reduction in TSB length, while offering smoother control strategies \cite{20}.

Despite considerable progress in flow control methods, the efficiency of CFC techniques has not been well documented. The existing literature primarily focuses on individual control methods, reinforcement learning algorithms, or specific applications of flow control actuators. However, a comprehensive comparison of these advanced methods, aimed at determining their relative effectiveness and potential for integration, has not been extensively explored. This study focuses on a systematic comparison of model predictive control (MPC) and RL-based approaches, including TDRL, DQL, and DQL integrated with signal processing, for active flow separation control.

\textcolor{black}{The motivation for this study stems from the need for more robust and efficient flow control strategies capable of maintaining high lift coefficients near stall conditions, where traditional control methods tend to lose effectiveness. While several studies have applied MPC or RL separately, few have systematically compared their performance in a challenging aerodynamic control scenario. The novelty of this work lies in the development and comparison of adaptive MPC and RL-based techniques for real-time, closed-loop control of a NACA 4412 airfoil under separated flow conditions, using a dual-point plasma actuator configuration initially introduced by the authors in earlier studies. Notably, the RL methods employed here are trained entirely online and within the simulation environment, without relying on pre-generated datasets. This distinguishes our approach from conventional data-driven strategies, which typically depend on offline training using fixed databases. This study provides valuable insight into control stability, performance trade-offs, and the adaptability of modern learning-based control approaches in aerodynamic applications.}

The rest of the article is structured as follows: the Flow Control Actuator section introduces the concept of dual-point excitation using plasma actuators. The underlying computational framework, including the governing equations, plasma actuator model, computational grid, and numerical methods used, are explained in the Computational Procedure section. In the Flow Control Algorithms section, the various algorithms are investigated: adaptive MPC, TDRL, DQL, and signal processing integrated with DQL, each described with their respective algorithms, governing equations, and code implementations. The Results and Discussion section presents a thorough analysis of the findings, highlighting the comparative performance of the different control strategies and discussing their implications for future research and practical applications. Finally, the Conclusions section provides a summary of the study’s key findings.

\section{Flow Control Actuator}
\textcolor{black}{
Active flow separation control can generally be classified into two principal approaches: direct and indirect. In the direct method, actuators such as wall jets~\cite{javadi2017separation, javadi2018quasi} inject high-momentum fluid directly into the near-wall region. This process energizes the boundary layer, enabling fluid particles to resist adverse pressure gradients more effectively, thus delaying or suppressing flow separation. The indirect approach, on the other hand, utilizes unsteady excitation of the separated shear layer, typically achieved through pulsating actuators. The fundamental mechanism involves the generation and propagation of disturbances and vortices which enhance the entrainment of high-momentum fluid into the separated region. This increased mixing accelerates the replacement of low-energy fluid near the wall with more energetic fluid from the free stream, ultimately contributing to the postponement of separation~\cite{samimy2018excitation, B100}.
}

\textcolor{black}{ Over the years, numerous actuator technologies have been investigated for applying periodic excitation to separated shear layers. Among these are synthetic jets, combustion-powered actuators, microfiber composite piezoelectric actuators, and plasma actuators. Notably, dielectric barrier discharge (DBD) plasma actuators have attracted growing interest due to several advantageous characteristics. These include their fully electric operation, absence of moving components, suitability for integration into low-profile surfaces like airfoils, rapid response to unsteady inputs, low weight, and relatively low power consumption~\cite{K1, oveisi2023experimental}. Despite these benefits, DBD plasma actuators also exhibit certain limitations. These include susceptibility to electromagnetic interference, low efficiency in converting electrical energy to mechanical momentum, and limited induced flow velocity, typically less than 8~m/s~\cite{louste2005sliding, moreau2007airflow}.}

\begin{figure}[h!]
    \centering
    \includegraphics[width=0.8\columnwidth]{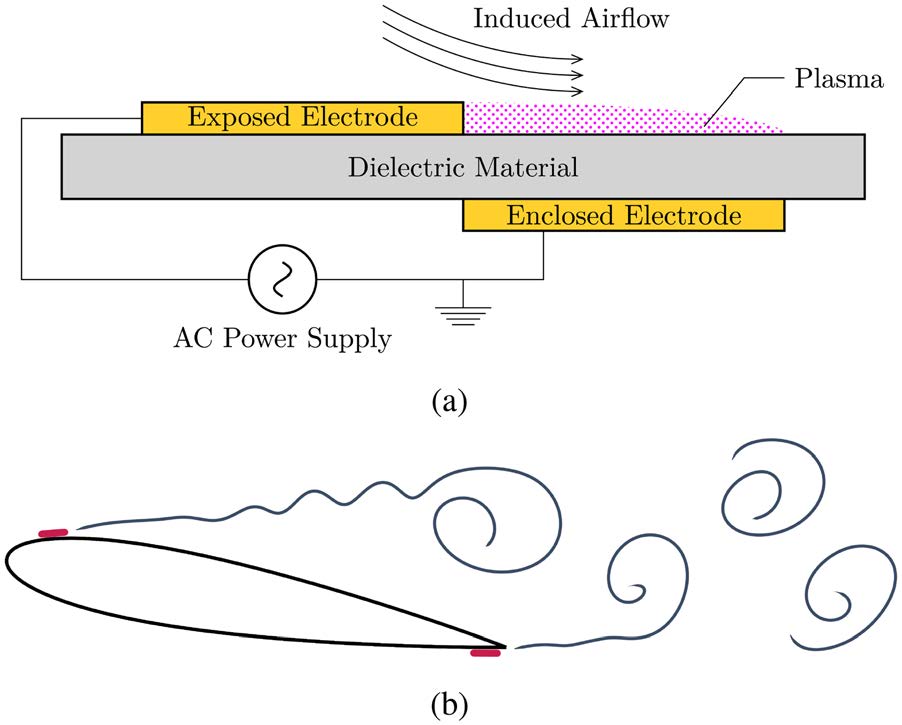}
    \caption{(a) Schematic of a DBD plasma actuator and (b) Schematic of dual-point excitation technique (actuators are indicated with red marks) \cite{B10}. \textcolor{black}{Source: Authors' own work.}}
    \label{fig:dbd_schematic}
\end{figure}

DBD plasma actuators consist of two electrodes: one that is exposed to the air and another that is covered by a dielectric material. As illustrated schematically in Fig.~\ref{fig:dbd_schematic}a, the system of a typical DBD plasma actuator operates by applying an AC voltage to the electrodes. When the voltage reaches a sufficiently high amplitude, it ionizes the air above the enclosed electrode. The ionized air particles are then accelerated by the electric field, exerting a body force on the surrounding neutral air. This interaction ultimately results in the generation of an induced airflow \cite{K3}.

In previous work by the authors \cite{ebrahimi2020dual}, the control of separated shear layers using DBD plasma actuators on a NACA 0015 airfoil was explored. The findings highlighted the actuator's success in stall prevention and its positive impact on lift-to-drag ratios. Li et al. \cite{10} employed plasma co-flow jets on a NACA 0025 airfoil, which significantly delayed flow separation through enhanced mixing shear layers.  Samimy et al. \cite{12} extended active flow control applications to high-speed flows using advancements in plasma actuators to manipulate Kelvin-Helmholtz instabilities. Additionally, Gross and Fasel \cite{16} studied separation control on a NACA 643-618 airfoil using plasma pulsed vortex generator jets, demonstrating high efficiency through linear stability analysis.

Controlling flow separation through plasma actuators, particularly by exciting the separated shear layer, is highly sensitive to the choice of excitation frequency. Research indicates that a free shear layer is most responsive to perturbations near its natural vortex shedding frequency, leading many studies to select frequencies within this range for optimal separation control \cite{B3, B5, B7}. However, some research favors higher excitation frequencies \cite{2, B1}. Another critical factor is the location of actuation, typically applied near the leading edge or the suction-side separation point to influence the upper separated shear layer \cite{B10}. Ebrahimi and Hajipour \cite{B100} introduced the concept of dual-point excitation, as shown in Fig.~\ref{fig:dbd_schematic}b, where both the suction side and trailing-edge shear layers are simultaneously excited using DBD plasma actuators. \textcolor{black}{By exploiting the interaction between the two shear layers, their method reduces the separation region and attenuates unsteady aerodynamic load fluctuations under stall conditions, demonstrating superior control performance compared to conventional single-point excitation methods. In this research, we apply the same dual-point excitation method using DBD plasma actuators as the foundation of the flow control strategy, while the appropriate excitation frequency is treated as a parameter to be determined by the control algorithms.}

\section{Computational Procedure}

\subsection{Governing equations}
\textcolor{black}{The numerical flow simulations are performed by analyzing a two-dimensional incompressible viscous Newtonian fluid with constant properties.} The governing equations employed for the Scale-Adaptive Simulation (SAS) turbulence model are based on the Reynolds-Averaged Navier-Stokes (RANS) equations, modified to include a scale-adaptive term that allows the model to transition between RANS and Large Eddy Simulation (LES) behavior based on the local flow conditions \cite{SAS1}. The continuity and momentum equations in Cartesian coordinates are:

\begin{equation}
\frac{\partial \rho}{\partial t} + \frac{\partial}{\partial x_i} (\rho u_i) = 0
\label{eq:1}
\end{equation}

\begin{equation}
\frac{\partial}{\partial t} (\rho u_i) + \frac{\partial}{\partial x_j} (\rho u_i u_j) = -\frac{\partial P}{\partial x_i} + \frac{\partial \tau_{ij}}{\partial x_j} + \rho f_i
\label{eq:2}
\end{equation}

Herein, $\rho$ is the fluid density, $u_i$ is the velocity component in the $i$-th direction, $P$ is the pressure, and $\tau_{ij}$ is the stress tensor which includes both molecular and turbulent contributions. The term $\rho f_i$ represents body forces, such as those due to plasma actuators. The stress tensor $\tau_{ij}$ is decomposed into the molecular viscosity part $\sigma_{ij}$ and the Reynolds stress part $\tau_{ij}^{T}$:

\begin{equation}
\sigma_{ij} = \mu \left( \frac{\partial u_i}{\partial x_j} + \frac{\partial u_j}{\partial x_i} \right) - \frac{2}{3} \mu \frac{\partial u_k}{\partial x_k} \delta_{ij}
\label{eq:3}
\end{equation}

\begin{equation}
\tau_{ij}^{T} = - \rho \overline{u_i' u_j'}
\label{eq:4}
\end{equation}

where $\mu$ is the dynamic viscosity of the fluid, and $\delta_{ij}$ is the Kronecker delta. The Reynolds stresses $\tau_{ij}^{T}$ are modeled using the SAS approach, which introduces an additional term to account for the resolved turbulent scales. In the SAS model, the additional source term $S_{SAS}$ is incorporated into the turbulence model equations, to adaptively scale the turbulence production and dissipation based on the local flow features. This term enables the model to transition smoothly between RANS and LES behaviors, capturing a wider range of turbulent scales without the need for fine-tuning.

In this study, the SAS model adapts dynamically to the resolved turbulence scales, providing a more accurate and computationally efficient representation of the flow dynamics compared to traditional RANS models. This approach ensures optimal performance in unsteady simulations, capturing essential flow features while reducing computational cost compared to full LES.

\subsection{Plasma actuator model}
In this study, the body force associated with plasma generation is derived from the phenomenological model introduced by Shyy et al. \cite{Plasma1}. This model strikes a balance between simplicity and precision, making it ideal for studies on plasma actuator-based flow control \cite{Plasma3, Plasma4}. Shyy et al. \cite{Plasma1} hypothesized that the plasma's electric field strength is linearly distributed within a triangular area directly downstream of the exposed electrode. Accordingly, the variation in field strength is calculated as:

\begin{equation}
|\vec{E}| = E_0 - k_1 x - k_2 y
\end{equation}

\noindent$E_0$ represents the maximum electric field strength within the triangular region, and the constants $k_1$ and $k_2$ are determined to ensure that the field strength at the boundary between the plasma and air (along the hypotenuse of the triangle) equals the breakdown value. Consequently, the body force within the plasma zone is given by:

\begin{equation}
\vec{F} = \rho_c e_c \vartheta \Delta T \vec{E}
\end{equation}

The field vector $\vec{E}$ is calculated as:

\begin{equation}
\vec{E} = \left(\frac{|\vec{E}| k_2}{\sqrt{k_1^2 + k_2^2}}, \frac{|\vec{E}| k_1}{\sqrt{k_1^2 + k_2^2}}, 0 \right)
\end{equation}

\noindent where, $\rho_c$ is the charge density of electrons, $e_c$ is the elementary charge, $\vartheta$ is the frequency of the applied voltage, and $\Delta T$ is the discharge time.

\begin{figure}[h!]
    \centering
    \includegraphics[width=0.8\columnwidth]{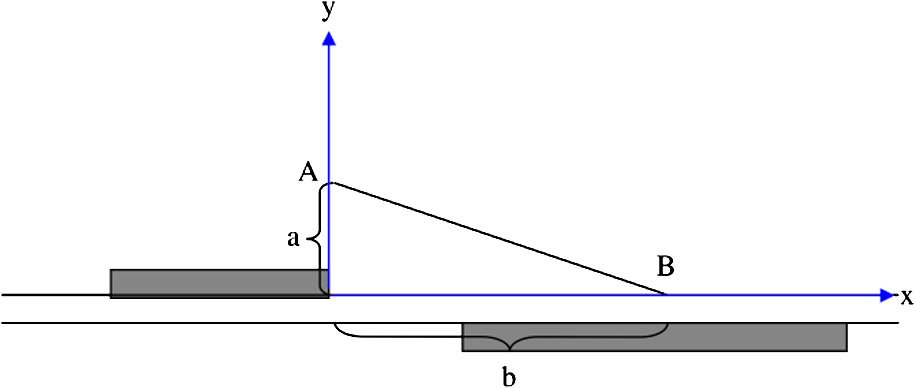}
    \caption{Triangular zone associated with plasma formation \cite{B100}. \textcolor{black}{Source: Authors' own work.}}
    \label{fig:dbd_2}
\end{figure}

In this research, the parameters of the plasma actuator are consistent with those used by Shyy et al. \cite{Plasma1}. The exposed and enclosed electrode lengths are 0.5 mm and 3 mm, respectively. The height of the electrodes is 0.1 mm, and the gap between them in the streamwise direction is 0.25 mm, as illustrated schematically in Fig.~\ref{fig:dbd_2}. The applied voltage is 9 kV (root mean square), with a frequency of 3 kHz. \textcolor{black}{Accordingly, the dimensions of the plasma actuation zone are specified as \( a = 0.0015\,\text{m} \) in height and \( b = 0.003\,\text{m} \) in length.} The numerical model of the DBD plasma actuator has been validated by comparing it with the work of Shy et al. \cite{Plasma1}. They assumed that plasma actuators operate in a steady, continuous mode; thus, the body force is modulated based on the duty cycle and excitation frequency to simulate unsteady actions. In this research, a duty cycle of 50\% and excitation frequencies ranging from \( f = 0 \) to 400 Hz, corresponding to a normalized frequency of \( F^+ = fc / U_{\infty} \) in the range of 0 to 11, are employed.

\subsection{Computational grid}

\begin{figure}[h!]
    \centering
    \includegraphics[width=0.8\columnwidth]{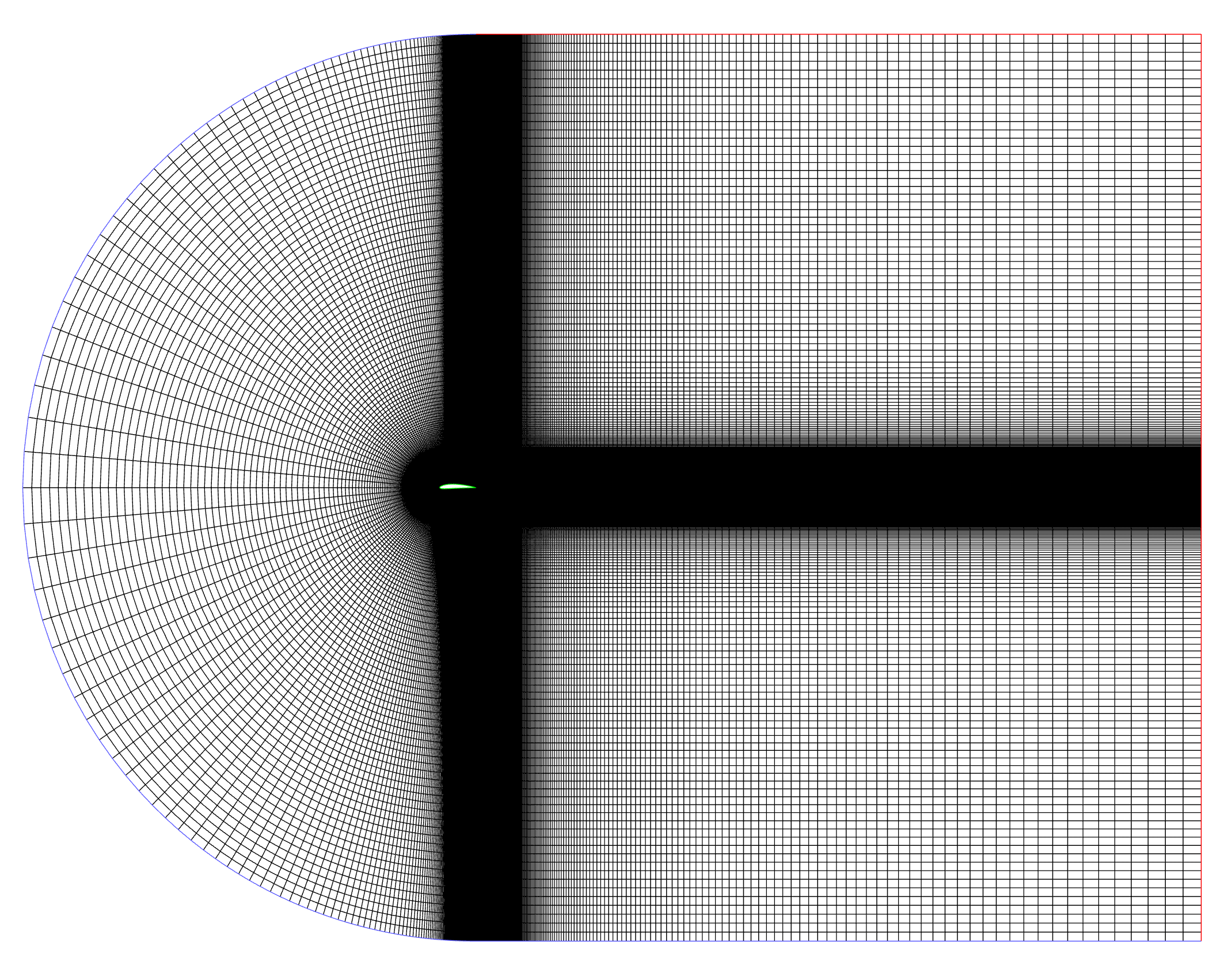}
    \caption{Side view of the computational domain. \textcolor{black}{Source: Authors' own work.}}
    \label{fig:400}
\end{figure}

The computational test case is based on a NACA 4412 airfoil with a chord length $c = 0.4 \, \text{m}$, $15^\circ$ angle of attack, and a Reynolds number $Re = 4.0 \times 10^5$. As depicted in Fig.~\ref{fig:400}, the computational domain extends $20c$ downstream, $12.5c$ upstream and $12.5c$ in the direction normal to the chord. The airfoil surface is modeled as an adiabatic wall with a no-slip boundary condition. A freestream with a constant velocity of 14.6 m/s enters the domain from the upstream and bottom boundaries, while the downstream and top boundaries are defined as pressure outlets with ambient pressure.

A structured C-type grid was employed to simulate the flow. The numerical grid consists of 471800 Quadrilateral cells, with 365 computational nodes on the airfoil's suction side and 280 nodes on the pressure side. The normal distance of the nearest node to the airfoil surface is $1 \times 10^{-5} \, \text{m}$, ensuring that the near-wall $y^+$ is kept below 1. To accurately predict separation, a dense grid resolution is used around the airfoil's trailing edge (i.e., the wake region).

\begin{figure}[h!]
    \centering
    \includegraphics[width=0.8\columnwidth]{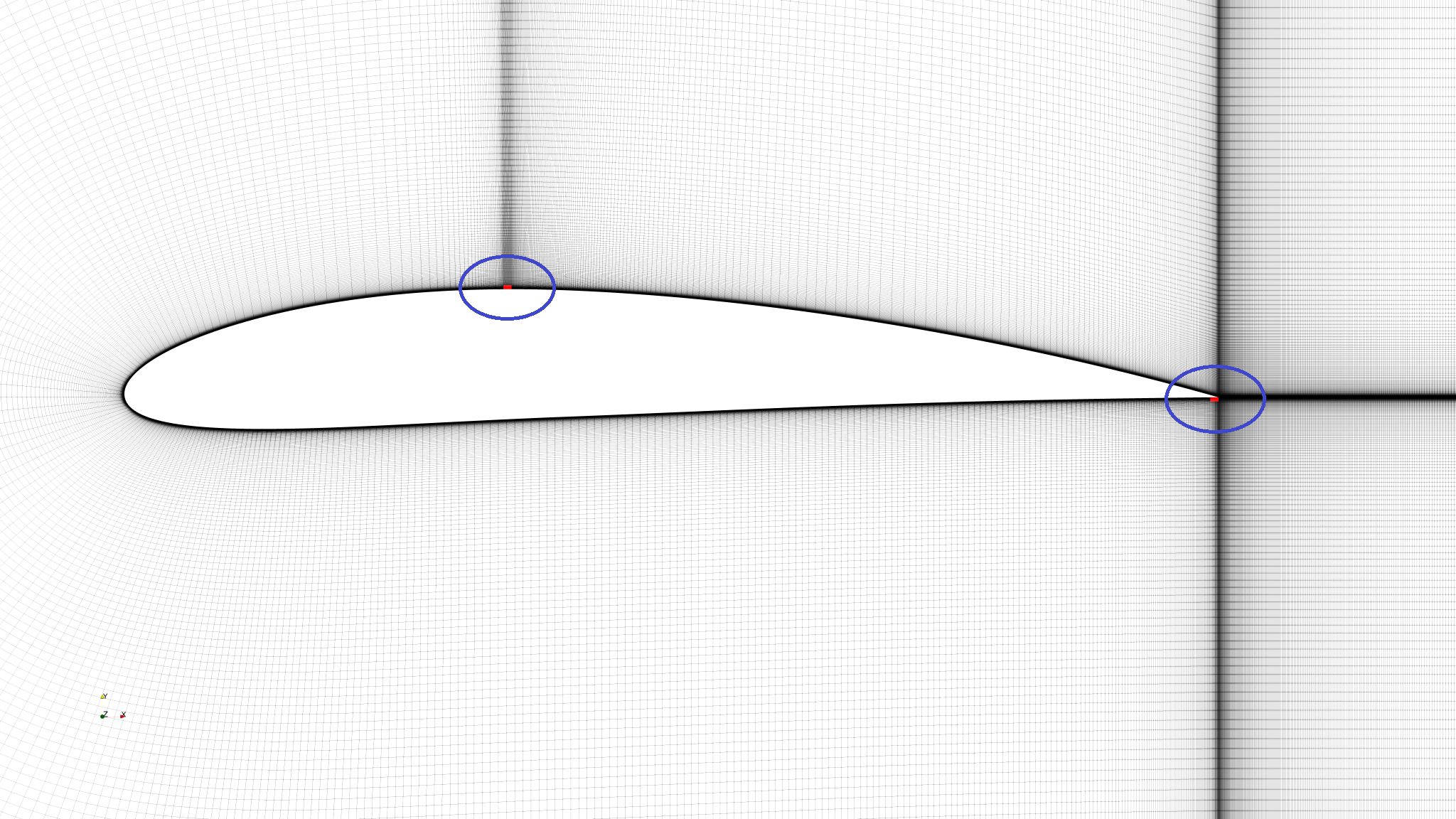}
    \caption{Detailed view of computational grid around airfoil and the plasma zones. \textcolor{black}{Source: Authors' own work.}}
    \label{fig:2}
\end{figure}

Following the previous work of authors \cite{B100}, the novel concept of dual excitation technique is utilized to control the separated shear layers. Two plasma zones are implemented in the grid: one on the suction side of the airfoil just downstream of the separated shear layer, and the other on the pressure side at the airfoil's trailing edge. The dimensions of these zones are selected based on the work by Shy et al. \cite{Plasma1}, with the plasma body force applied only within these regions. The suction side actuator is located at $x = 0.35c$, while the pressure side actuator is positioned at $x = 0.99c$. The grid's streamwise resolution in the plasma zones is chosen carefully to satisfy the accuracy requirements for predicting plasma wall jets and flow separation. Fig.~\ref{fig:2} illustrates a detailed view of the plasma zones encircled in blue and the computational grid around the airfoil.

\subsection{Numerical method}
The computational fluid dynamics (CFD) simulations in this study were conducted using a pressure-based finite volume solver. The simulations were transient and performed in a two-dimensional planar space, with gravity effects neglected. In this study, air was modeled with constant properties, specifically, a density of $1.225$~kg/m$^{3}$ and a dynamic viscosity of $1.7894 \times10^{-5}$~kg/(m$\cdot$s). \textcolor{black}{The assumption of constant properties is a common simplification in low-speed aerodynamic simulations, where variations in temperature and pressure have a negligible effect on fluid properties. DBD plasma actuators are typically categorized as cold-plasma devices, meaning that plasma generation does not involve significant thermal input. Accordingly, the temperature variations induced by plasma are highly localized and do not meaningfully influence the bulk flow characteristics, particularly under the low Mach number and 50\% duty cycle conditions applied in this study. This approach aligns with established assumptions in similar validated CFD studies \cite{B100, B1, Plasma1}} 

A coupled scheme with Rhie-Chow momentum-based flux type \cite{RCNS} was implemented for pressure and velocity fields. To achieve high accuracy in the numerical solution, least squares cell-based gradient discretization was applied. Second-order discretization was used for pressure, while bounded central differencing was applied for momentum. Second-order upwind schemes were used for both turbulent kinetic energy and specific dissipation rate. The transient nature of the flow was captured using a bounded second-order implicit scheme, which provides a balance between stability and accuracy for time-dependent simulations. \textcolor{black}{The CFD solver time step was set to $10^{-4}$~s, which is sufficiently smaller than the excitation periods associated with the tested actuation frequencies (50–400 Hz). This choice ensures adequate temporal resolution to accurately capture the unsteady effects of plasma forcing. The numerical setup described in this section was carefully selected to ensure accurate and robust simulation of the flow dynamics under active flow control.}

\section{Flow Control Algorithms}

The efficiency of CFC techniques is significantly influenced by the underlying algorithms, as they directly impact the ability to optimize fluid flow behavior and achieve flow control objectives. In order to conduct a comparative analysis, this study evaluates Adaptive MPC and RL-based techniques based on their ability to optimize excitation frequency and expedite the identification of stable conditions. Approaches like MPC rely on deriving a dynamic model between input and output to predict and adjust flow parameters in real time, offering precise control based on the identified model accuracy. Recently, RL techniques such as temporal difference (TD) and DQL have emerged as powerful alternatives, capable of learning optimal control policies through environmental interaction and feedback, thus providing model-free solutions. Furthermore, integrating signal processing with DQL provides advanced feature extraction from complex flow data, resulting in more effective and precise control strategies. This section delves into the various flow control algorithms, highlighting their underlying principles, advantages, and applications in fluid dynamics.\\
All the methods evaluated in this study are implemented in an online learning manner, where each technique trains while simultaneously providing outputs for the simulation, enabling real-time adaptation. For the RL-based methods, hyperparameters were determined after multiple initial runs. The selection of hyperparameters was based on the fact that the model initially explores its options and selects the optimal action as quickly as possible. In each model, the lift coefficient serves as the input, while the output is the actuation frequency. The action space for the RL-based methods is designed to be discrete, as the desired frequency value for flow control lies within the superharmonic frequencies of wake, specifically from $F^+=1$ to $F^+=6$. Frequencies beyond this range do not elicit an effective fluid response to excitation \cite{2}. Moreover, the online learning approach involves significant computational demands, emphasizing the importance of using efficient, lightweight models to balance computational feasibility and control effectiveness. The authors explored more complex network architectures beyond those presented in this paper and found that increasing complexity led to a decline in performance.
\subsection{Adaptive MPC}
MPCs are renowned for their ability to control complex processes with great performance in energy optimization, robustness, stability, and disturbance rejection \cite{S1, S2, S3, S4}. This class of controllers requires a model to predict future plant outputs and then take actions to minimize future errors between desired values and actual outputs. In this section, we aim to design an adaptive Generalized Predictive Controller (GPC) that can fit a linear model between the control input \( u(t) \) (i.e., actuator excitation frequency) and the plant output \( y(t) \) (i,e., airfoil lift coefficient) using the Recursive Least Squares (RLS) system identification method.

The basic idea of GPC is to calculate a sequence of future control inputs in such a way that it minimizes a multistage cost function defined over a prediction horizon 
 \cite{S5}. To do this, we first define a Controller Auto-Regressive Integrated Moving-Average (CARIMA or ARIMAX) model \cite{S6} for describing the GPC algorithm:

\begin{equation}
A(z^{-1})y(t) = B(z^{-1})u(t-1) + C(z^{-1})\frac{e(t)}{\Delta} \quad \text{with} \quad \Delta = 1 - z^{-1}
\label{eq:11}
\end{equation}

where \( z = e^{sT_c} \) is a discrete-time equivalent of the Laplace variable \( s \), and \( T_c \) is the controller time step. Also, \( e(t) \) is zero-mean white noise and \( A, B, C \) are polynomials in the backward shift operator \( z^{-1} \):

\begin{equation}
\begin{aligned}
A(z^{-1}) &= 1 + a_1z^{-1} + a_2z^{-2} + \cdots + a_{n_a}z^{-n_a} \\
B(z^{-1}) &= b_0 + b_1z^{-1} + b_2z^{-2} + \cdots + b_{n_b}z^{-n_b} \\
C(z^{-1}) &= 1 + c_1z^{-1} + c_2z^{-2} + \cdots + c_{n_c}z^{-n_c}
\end{aligned}
\label{eq:12}
\end{equation}

In this study, we consider \( C(z^{-1}) = 1 \), meaning we have only white noise in our process. Due to the complexity of the problem, a system identification method is needed to estimate the \( A \) and \( B \) polynomials discussed in the system identification section. The GPC algorithm consists of applying a control sequence that minimizes a multistage cost function of the form:

\begin{equation}
J(N_p, N_u) = \sum_{j=1}^{N_p} \left( \hat{y}(t+j|t) - w(t+j) \right)^2 + \sum_{j=1}^{N_u} \lambda(j) \left( \Delta u(t+j-1) \right)^2
\label{eq:13}
\end{equation}

\noindent where \( N_p \) is the maximum costing horizon, also known as the prediction horizon, and \( N_u \) is the control horizon. \( \lambda(j) \) is a control-weighting sequence. Additionally, \( \hat{y}(t + j|t) \) is an optimum \( j \)-step ahead prediction of the system output based on data up to time \( t \) and \( w(t + j) \) is the future reference trajectory. This implies that in the GPC algorithm, an achievable desired value should be defined for the lift coefficient and used as \( w(t) \) in the cost function. To derive the \( j \)-step ahead prediction of the system output \( y(t + j) \) based on Eq.~\eqref{eq:11}, consider the following Diophantine equation:

\begin{equation}
1 = E_j(z^{-1})\tilde{A}(z^{-1}) + z^{-j}F_j(z^{-1}) \quad \text{with} \quad \tilde{A}(z^{-1}) = \Delta A(z^{-1})
\label{eq:14}
\end{equation}

\noindent where polynomials \( E_j \) and \( F_j \) are uniquely defined for a given \( \tilde{A} \) that have degrees of \( j - 1 \) and \( n_a \), respectively. If Eq.~\eqref{eq:11} is multiplied by \( \Delta E_j(z^{-1})z^j \) and using Eq.~\eqref{eq:14}, we can find the \( j \)-step ahead prediction of the system output:

\begin{equation}
y(t + j) = G_j(z^{-1})\Delta u(t + j - 1) + F_j(z^{-1})y(t) + E_j(z^{-1})e(t + j)
\label{eq:15}
\end{equation}

\noindent where \( G_j(z^{-1}) = E_j(z^{-1})B(z^{-1}) \). Thus, the degree of polynomial \( E_j(z^{-1}) \) equals \( j - 1 \), the noise term \( e(t + j) \) in equation Eq.~\eqref{eq:15} is all in the future. Therefore the best prediction of \( y(t + j) \) is:

\begin{equation}
\hat{y}(t + j|t) = G_j(z^{-1})\Delta u(t + j - 1) + F_j(z^{-1})y(t)
\label{eq:16}
\end{equation}

Following from Ref. \cite{S6}, it is simple to show that the required polynomials \( E_j \) and \( F_j \) in Eq.~\eqref{eq:16} can be obtained recursively as:

\begin{align}
E_j(z^{-1}) &= e_{j,0} + e_{j,1}z^{-1} + \cdots + e_{j,j-1}z^{-(j-1)} \nonumber \\
F_j(z^{-1}) &= f_{j,0} + f_{j,1}z^{-1} + \cdots + f_{j,n_a}z^{-n_a}
\label{eq:17}
\end{align}

\noindent following the recursion algorithm:

\begin{equation}
\begin{aligned}
&E_1 = 1; \quad F_1 = z \left( 1 - \tilde{A}(z^{-1}) \right) \\
&E_{j+1}(z^{-1}) = E_j(z^{-1}) + f_{j,0}z^{-j} \\
&f_{j+i} = f_{j,i+1} - f_{j,0}\tilde{a}_{i+1} \quad (i = 0, 1, \ldots, n_a - 1)
\end{aligned}
\label{eq:18}
\end{equation}

\noindent also, \( G_{j+1} \) can be calculated recursively:
\begin{equation}
G_{j+1} = E_{j+1}B = \left( E_j + f_{j,0}z^{-j} \right)B = G_j + f_{j,0}z^{-j}B
\label{eq:19}
\end{equation}
Considering Eq.~\eqref{eq:16}, we can write a set of \( j \)-ahead predictions:

\begin{equation}
\begin{aligned}
\hat{y}(t + 1|t) &= G_1 \Delta u(t) + F_1 y(t) \\
\hat{y}(t + 2|t) &= G_2 \Delta u(t + 1) + F_2 y(t) \\
&\vdots \\
\hat{y}(t + N_p|t) &= G_{N_p} \Delta u(t + N_p - 1) + F_{N_p} y(t)
\end{aligned}
\label{eq:110}
\end{equation}

\noindent which can be displayed in the following matrix form:

\begin{equation}
\hat{\mathbf{y}} = \mathbf{G} \mathbf{u} + \mathbf{f} \label{eq:111}
\end{equation}

This form shows that the response of the system consists of two parts: forced response \( \mathbf{G} \mathbf{u} \) and free response \( \mathbf{f} \). Matrix \( \mathbf{G} \) and vectors \( \hat{\mathbf{y}}, \mathbf{u} \), and \( \mathbf{f} \) in Eq.~\eqref{eq:111} can be defined as follows:

\begin{equation}
\begin{aligned}
&\mathbf{G} = \begin{bmatrix}
g_0 & 0 & \cdots & 0 \\ g_1 & g_0 & \cdots & 0 \\ \vdots & \vdots & \ddots & \vdots \\ \vdots & \vdots & \cdots & g_0 \\ \vdots & \vdots & \ddots & \vdots \\ g_{N_p-1} & g_{N_p-2} & \cdots & g_{N_p-N_u} \end{bmatrix} \\
&\hat{\mathbf{y}} = \begin{bmatrix}
\hat{y}(t + 1|t), & \hat{y}(t + 2|t), & \cdots, & \hat{y}(t + N_p|t)
\end{bmatrix}^T \\
&\mathbf{u} = \begin{bmatrix}
\Delta u(t), & \Delta u(t + 1), & \cdots, & \Delta u(t + N_u - 1)
\end{bmatrix}^T \\
&\mathbf{f} = \begin{bmatrix}
f(t + 1), & f(t + 2), & \cdots, & f(t + N_p)
\end{bmatrix}^T\\
\end{aligned}
\label{eq:112}
\end{equation}

\noindent with \(f(t + i) = \left( G_i(z^{-1}) - \sum_{k=0}^{i-1} g_k z^{-k} \right) z^i \Delta u(t) + F_i(z^{-1}) y(t)\) for \(i = 1, \ldots, N_p\) and considering \( N_p \geq N_u \), which means that \( \Delta u(t + i) = 0 \) for \( i \geq N_u \). Therefore, the cost function provided in Eq.~\eqref{eq:13} can be rewritten in the following form:

\begin{equation}
\begin{aligned}
&J = (\mathbf{G} \mathbf{u} + \mathbf{f} - \mathbf{w})^T (\mathbf{G} \mathbf{u} + \mathbf{f} - \mathbf{w}) + \lambda \mathbf{u}^T \mathbf{u}\\
&\mathbf{w} = \begin{bmatrix}
w(t + 1), & w(t + 2), & \cdots, & w(t + N_p)
\end{bmatrix}^T
\end{aligned}
\label{eq:113}
\end{equation}

Assuming there are no constraints on future controls, the minimum of \( J \) can be found by making \( \frac{\partial J}{\partial \mathbf{u}} = 0 \), which leads to:

\begin{equation}
\mathbf{u} = (\mathbf{G}^T \mathbf{G} + \lambda I)^{-1} \mathbf{G}^T (\mathbf{w} - \mathbf{f})
\label{eq:114}
\end{equation}

A system identification method is needed that can estimate polynomials \( A \) and \( B \) of the system to describe control input \( u(t) \) and complete Adaptive GPC design. According to equation Eq.~\eqref{eq:11}, we can perform the RLS method \cite{S7} as follows:

\begin{equation}
\begin{aligned}
&k(t) = P(t - 1) \phi(t) \left( I + \phi^T(t) P(t - 1) \phi(t) \right)^{-1}\\
&P(t) = \left( I - k(t) \phi^T(t) \right) P(t - 1) \\
&\theta(t) = \theta(t - 1) + k(t) \left( y(t) - \phi^T(t) \theta(t - 1) \right)\\
\end{aligned}
\label{eq:116}
\end{equation}

\noindent where the regressors vector \( \phi(t) \) and parameters vector \( \theta(t) \) are defined as follows:

\begin{equation}
\begin{aligned}
&\phi(t) = \left[ -y(t - 1) \enspace \cdots \enspace -y(t - n_a) \enspace u(t - 1) \enspace \cdots \enspace u(t - n_b) \right]^T \\
&\theta(t) = \left[ a_1 \enspace \cdots \enspace a_{n_a} \enspace b_1 \enspace \cdots \enspace b_{n_b} \right]^T
\end{aligned}
\label{eq:117}
\end{equation}

The system output \(y(t)\) is the average of the lift coefficients from the last \(\frac{\text{FoM} \times T_c}{\text{TimeStep}}\) intervals, calculated over each \(T_c\), where \(\text{FoM}\) stands for Fraction of Mean and is set between 0 and 1.

The parameters required for implementing the Adaptive GPC algorithm are described in Table~\ref{table:parameters_mpc}. These parameters are used in the results.
\begin{table}[h!]
\centering
\caption{Parameter values for adaptive GPC algorithm. \textcolor{black}{Source: Authors' own work.}}
\begin{tabular}{lcl}
\hline
\textbf{Parameter} & \textbf{Value} & \textbf{Description} \\ \hline
Input Range        & [50, 400]      & Range of the continuous control inputs (Hz)      \\
FoM                & 0.25           & Fraction of mean      \\
\(\lambda\)        & 1              & Control weight        \\
\(n_a\)            & 5              & Order of polynomial A \\
\(n_b\)            & 5              & Order of polynomial B \\
\(N_p\)            & 11             & Prediction horizon    \\
\(N_u\)            & 7              & Control horizon       \\
\(T_c\)            & 0.1            & Controller time step (s)  \\
TimeStep           & 1e-4           & Simulation time step (s)  \\ \hline
\end{tabular}
\label{table:parameters_mpc}
\end{table}

\subsection{TDRL}
TDRL is a class of model-free methods used in reinforcement learning, where an agent learns to make decisions by interacting with an environment. Unlike traditional Monte Carlo methods that wait until the end of an episode to update value estimates, TD methods update estimates based on partially observed episodes, making them more efficient for online learning. TD learning combines ideas from Monte Carlo methods and dynamic programming.\\
The core idea behind TD learning is to estimate the value function, which represents the expected return (future rewards) for being in a given state and following a particular policy. The TD learning algorithm iteratively updates the value estimates using the difference between the predicted value and the actual reward received, known as the TD error.\\
The reward is calculated based on the difference between the current mean lift coefficient ($C_{l_{\text{mean}}}$), which is the mean of the last \(\frac{\text{FoM} \times T_c}{\text{TimeStep}}\) lift coefficients in a Controller Time Step, and a baseline value, using an exponential function to emphasize larger improvements. The current state is assumed to be the previous reward. The Q-value for the selected action is updated as Eq.~\eqref{eq:21} using the TD error, incorporating the immediate reward and the maximum future Q-values.
\begin{equation}
Q(s, a) \leftarrow Q(s, a) + \alpha \left(r + \gamma \max_{a'} Q(s', a') - Q(s, a) \right)
\label{eq:21}
\end{equation}
where, \( s \) is the current state from which the action \( a \) is taken, \( s' \) is the new state resulting from taking action \( a \), \( \alpha \) is the learning rate, \( r \) is the reward received after taking action \( a \) in state \( s \), \( \gamma \) is the discount factor, \( Q(s, a) \) is the current Q-value for taking action \( a \) in state \( s \), and \( \max_{a'} Q(s', a') \) is the maximum Q-value among all possible actions from the new state \( s' \).
The algorithm decides whether to explore new actions or exploit the current knowledge by comparing a random number with the exploration rate ($\epsilon$). $\epsilon$ is decayed by multiplying it by exploration decay ($\kappa$) in every controller time step. The parameters required for implementing the TDRL algorithm are described in Table~\ref{table:combined_parameters}.

\begin{table}[ht]
\centering
\caption{Parameter values for TD and DQL algorithms. \textcolor{black}{Source: Authors' own work.}}
\begin{tabular}{lcll}
\hline
\textbf{Parameter}             & \textbf{TDRL value}  & \textbf{DQL value} & \textbf{Description}                    \\ \hline
Actions                        & 0:50:400           & 0:50:400           & Discrete action space      \\
$\alpha$                       & 0.4                & \( 1 \times 10^{-2} \)               & Learning rate                           \\
BaseLine                       & 1.44               & 1.44               & Baseline lift coefficient               \\
$\epsilon$                     & 0.9                & 0.9                & Initial exploration rate                \\
FoM                            & 0.25               & 0.25               & Fraction of mean                        \\
$\gamma$                       & 0.9                & 0.25               & Discount factor                         \\
$\kappa$                       & 0.95               & 0.95               & Exploration rate decay       \\
\(T_c\)                        & 0.1                & 0.1                & Controller time step                    \\
TimeStep                       & \( 1 \times 10^{-4} \)               & \( 1 \times 10^{-4} \)               & Simulation time step                    \\ \hline
\end{tabular}
\label{table:combined_parameters}
\end{table}

\subsection{Deep Q - Learning}
DQL is a model-free RL algorithm that combines Q-Learning with deep neural networks. It is designed to handle environments with high-dimensional state spaces where traditional Q-Learning would be infeasible \cite{mnih2015human}. DQL approximates the Q-value function using a neural network, which takes the current state as input and outputs Q-values for each possible action. The algorithm iteratively updates the Q-values based on the observed rewards and transitions. Like the previous method, rewards and states are calculated based on the environment's responses to the agent's actions. The Q-value update rule is given by Eq.~\eqref{eq:21}. \textcolor{black}{Where $\alpha$ is the learning rate that determines the extent to which the new information overrides the old information. If $\alpha = 1$, the equation simplifies to:}

\begin{equation}
Q(s,a) \leftarrow r + \gamma \max_{a'} Q(s', a')
\end{equation}

where, $Q(s, a)$ is the Q-value for taking action $a$ in state $s$, $r$ is the reward received after taking action $a$, $\underset{a'}{\max} Q(s',a')$ is the maximum predicted Q-value for the next state $s'$, considering all possible actions $a'$, and $Q(s', a')$ is the Q-value function approximator (neural network) used for predicting future rewards.\\
The \(\epsilon\)-greedy policy is used to balance exploration and exploitation. The neural network used in DQL is designed to approximate the Q-value function. It takes the current state as input and outputs the Q-values for all possible actions. The architecture of the network in this implementation is as follows:
\begin{itemize}
    \item \textbf{Input Layer}: A feature input layer with 1 input feature.
    \item \textbf{Hidden Layers}: Four fully connected layers, each with 4 neurons and ReLU activation functions. These layers allow the network to learn complex representations of the state-action space.
    \item \textbf{Output Layer}: A fully connected layer with the number of neurons equal to the number of possible actions. This layer outputs the Q-values for each action.
\end{itemize}

\begin{figure}[h!]
    \centering
    \includegraphics[width=\columnwidth]{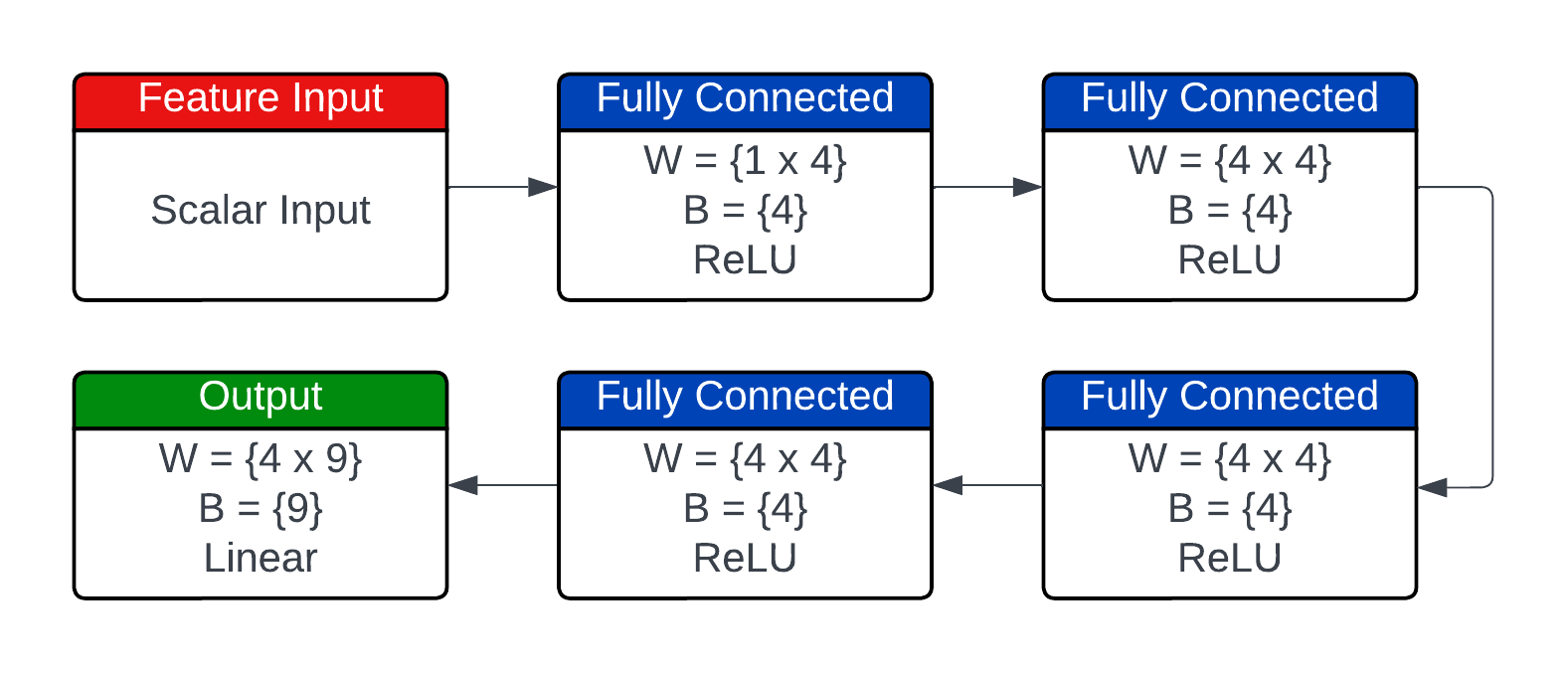}
    \caption{DQL Network Architecture. W and B indicate weight and Bias of layers, respectively. \textcolor{black}{Source: Authors' own work.}}
    \label{fig:qvalue_graph}
\end{figure}

The network is trained using the Stochastic Gradient Descent with Momentum (SGDM) optimizer, with an initial learning rate of 0.01 and a maximum of 100 epochs, using half-mean-squared error as the loss function. This setup allows the network to learn the optimal Q-value function through iterative updates based on the observed rewards and transitions. The network architecture and the parameters required for implementing the DQL algorithm are described in Fig.~\ref{fig:qvalue_graph} and Table~\ref{table:combined_parameters}, respectively.

\subsection{Signal processing integrated with Deep Q-Learning}
Signal processing integrated with DQN aims to enhance the algorithm's ability to handle and interpret time-series data effectively. By incorporating Long Short-Term Memory (LSTM) \cite{schmidhuber1997long} layers within the neural network architecture, the algorithm can better understand temporal dependencies and patterns in the data, leading to more informed decision-making. This approach is particularly useful in environments where the state information is derived from sequences of observations, such as time-series signals. The following describe the algorithm and the governing equations used in this integrated approach.\\
The neural network in this approach takes sequences of states as input and outputs Q-values for each possible action. The algorithm iteratively updates the Q-values based on the observed rewards and transitions, allowing the agent to learn optimal actions over time by recognizing patterns and dependencies in the sequential data. This integration of signal processing enhances the agent's decision-making ability in dynamic environments.\\
As in the previous method, rewards and states are calculated based on the environment's responses to the agent's actions. The primary difference lies in the state representation, which is now a vector comprising the last \(\frac{\text{FoM} \times T_c}{\text{TimeStep}}\) of lift coefficients over each controller time step. Despite this change in state and reward definitions, the governing equations, methodology, and \(\epsilon\)-greedy policy and network training options remain identical to those used in the prior approach. The parameters are identical to the previous method.\\
\begin{figure}[h!]
    \centering
    \includegraphics[width=\columnwidth]{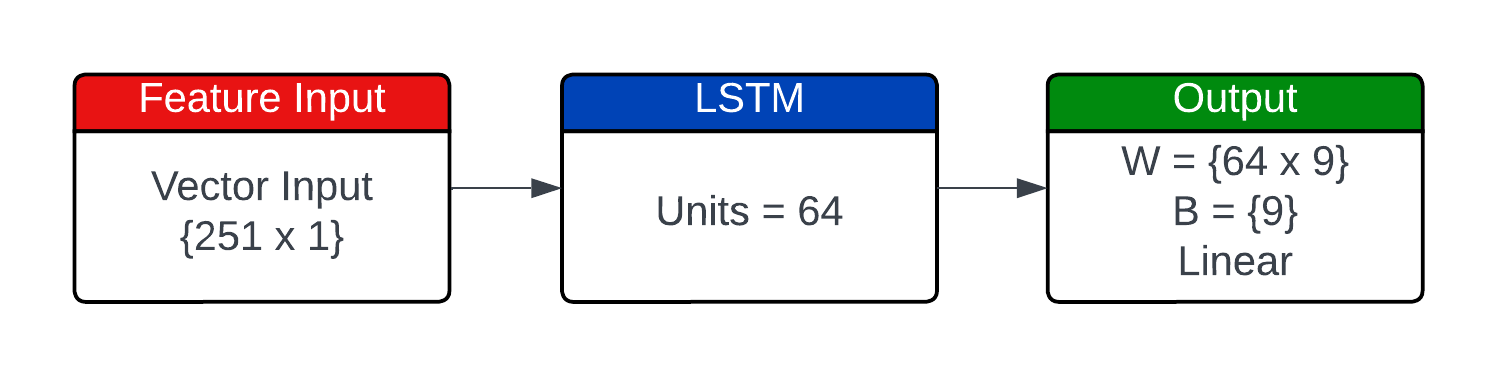}
    \caption{Signal processing integrated with DQL Network Architecture. \textcolor{black}{Source: Authors' own work.}}
    \label{fig:DQNSP}
\end{figure}
The neural network used in this implementation of Deep Q-Learning is specifically designed to handle time-series data. It takes sequences of states as input and outputs the Q-values for all possible actions. The architecture of the network is structured as follows (see Fig.~\ref{fig:DQNSP}):

\begin{itemize}
    \item \textbf{Input Layer}: A feature input layer with 251 input features.
    \item \textbf{LSTM Layer}: An LSTM (Long Short-Term Memory) layer with 64 neurons, designed to capture temporal dependencies in the data.
    \item \textbf{Fully Connected Layer}: A fully connected layer with the number of neurons equal to the number of possible actions, outputting the Q-values for each action.
\end{itemize}

\section{Results and Discussion}

This section presents the outcomes of different control strategies applied to the airfoil flow separation problem. The section begins with validating the baseline model using experimental results from wind tunnel tests of Mallor et al. \cite{mallor2024experimental} and Tabatabaei et al. \cite{refexp}. After the baseline is established, the performance of adaptive MPC is analyzed and these results are compared with those obtained from Temporal Difference Reinforcement Learning and Deep Q-learning approaches. The effectiveness of integrating signal processing techniques with Deep Q-learning is also explored. The discussion focuses on evaluating control performance, stability, and computational efficiency across these methodologies.

\subsection{Baseline Flow Simulation and Validation}

\begin{figure}[h!]
    \centering
    \includegraphics[width=0.8\columnwidth]{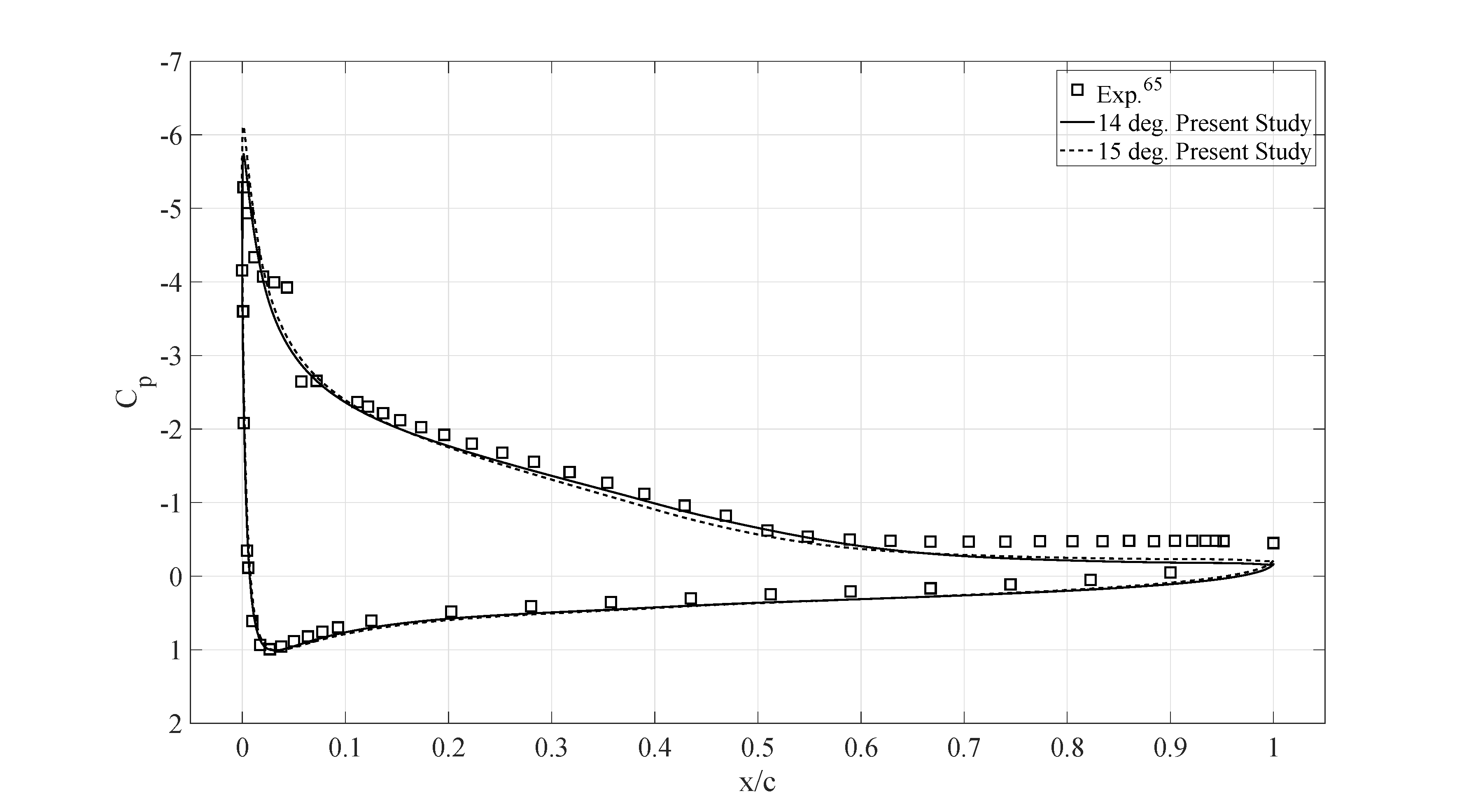}
    \caption{Comparison of \(C_p\) distribution along the \(x/c\) between experimental and CFD data. \textcolor{black}{Source: Authors' own work.}}
    \label{fig:Cpvsx}
\end{figure}

\begin{figure}[h!]
    \centering
    \includegraphics[width=0.8\columnwidth]{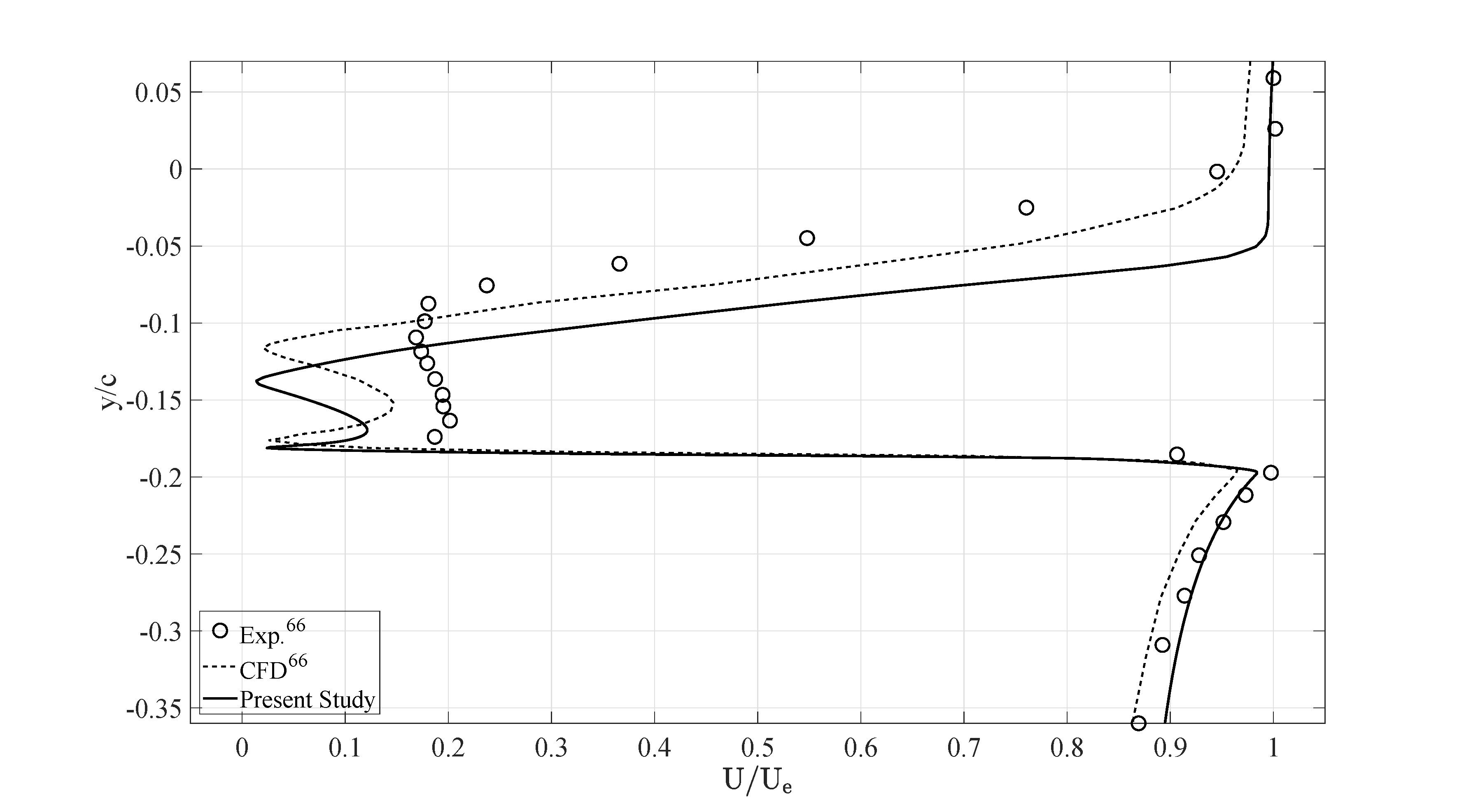}
    \caption{Comparison of wake velocity profiles at the trailing edge. \textcolor{black}{Source: Authors' own work.}}
    \label{fig:Wake}
\end{figure}

This subsection focuses on the baseline flow simulation and its validation against experimental data. Fig.~\ref{fig:Cpvsx} presents the pressure coefficient (\( C_p \)) distribution along the airfoil surface, compared with the experimental data of Mallor et al.~\cite{mallor2024experimental} at \( 14^\circ \) angle of attack and Reynolds number \( Re = 4.0 \times 10^5 \), showing satisfactory agreement. Furthermore, the dimensionless wake velocity profile at the trailing edge is compared with the experimental and numerical data of Tabatabaei et al.~\cite{refexp} in Fig.~\ref{fig:Wake}, where \( U_e \) denotes the edge velocity of the wake. The results indicate that the present simulation captures the overall trend observed in both experimental and numerical data. As this study aims to investigate near-stall conditions, the remainder of the paper focuses on \( \alpha = 15^\circ \).

\begin{figure}[h!]
    \centering
    \begin{subfigure}[b]{0.48\columnwidth}
        \centering
        \includegraphics[width=\linewidth]{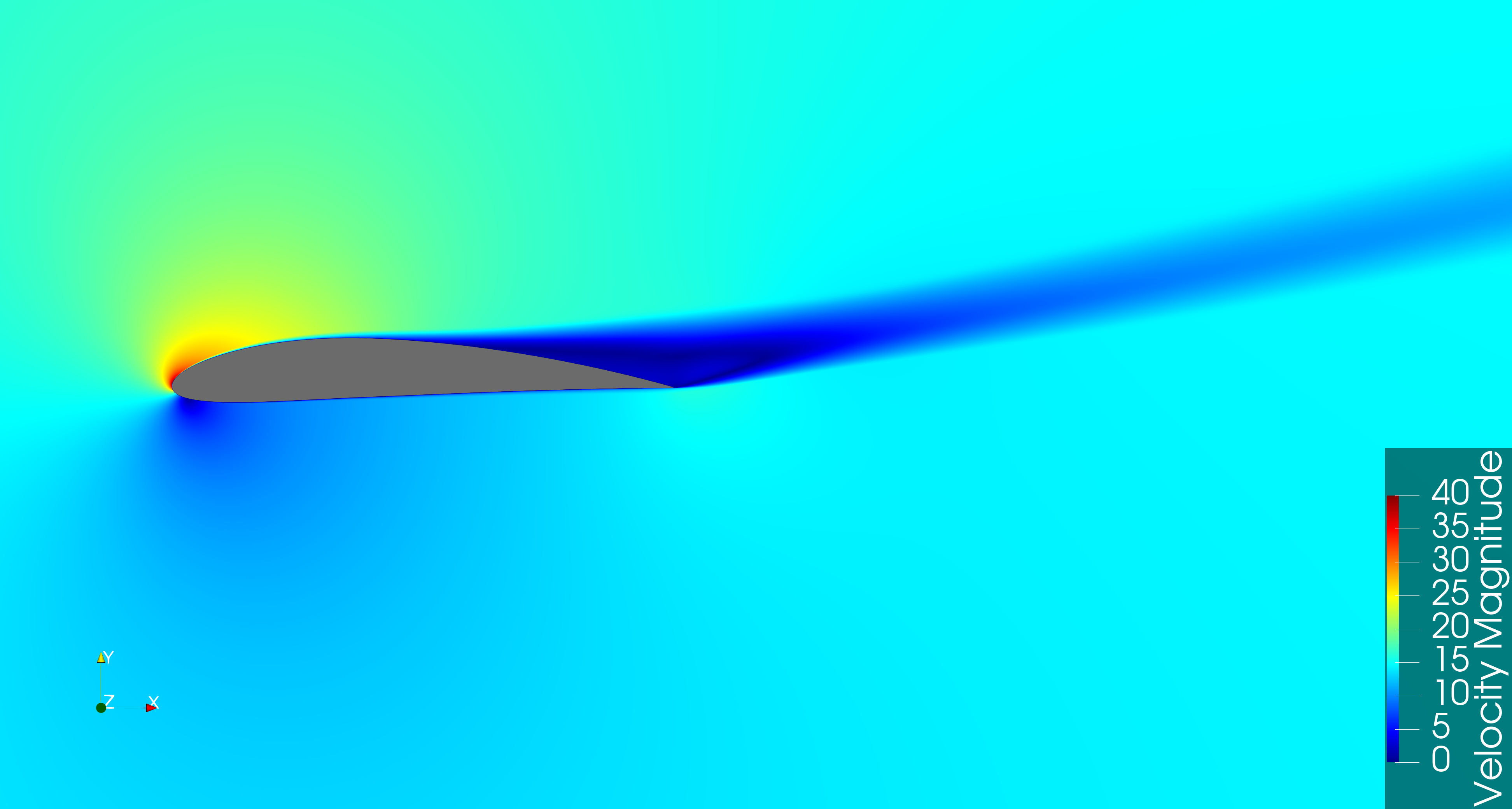}
        \caption{}
        \label{fig:VMCBF}
    \end{subfigure}
    \hfill
    \begin{subfigure}[b]{0.48\columnwidth}
        \centering
        \includegraphics[width=\linewidth]{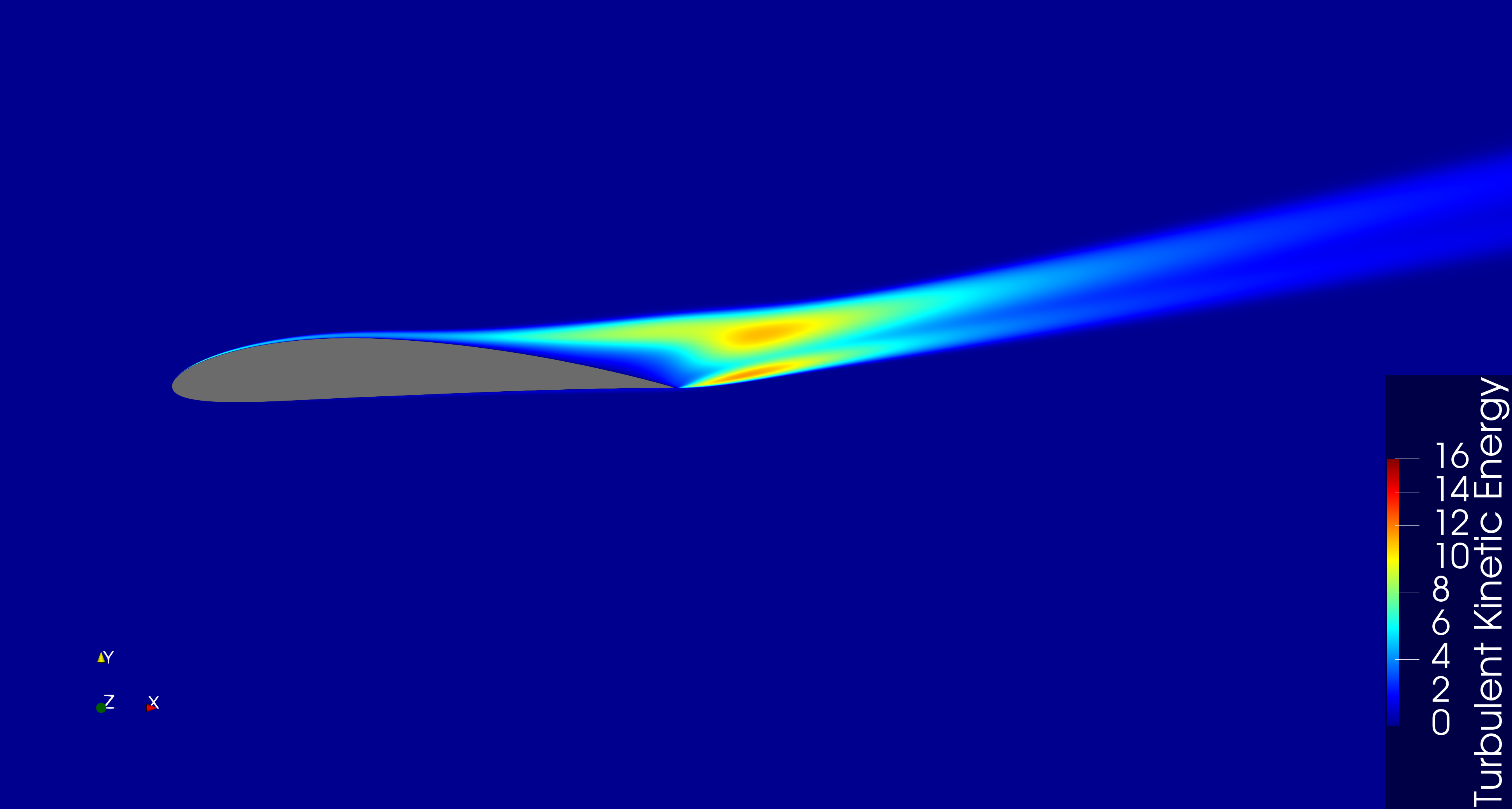}
        \caption{}
        \label{fig:TKECBF}
    \end{subfigure}
    
    \begin{subfigure}[b]{0.48\columnwidth}
        \centering
        \includegraphics[width=\linewidth]{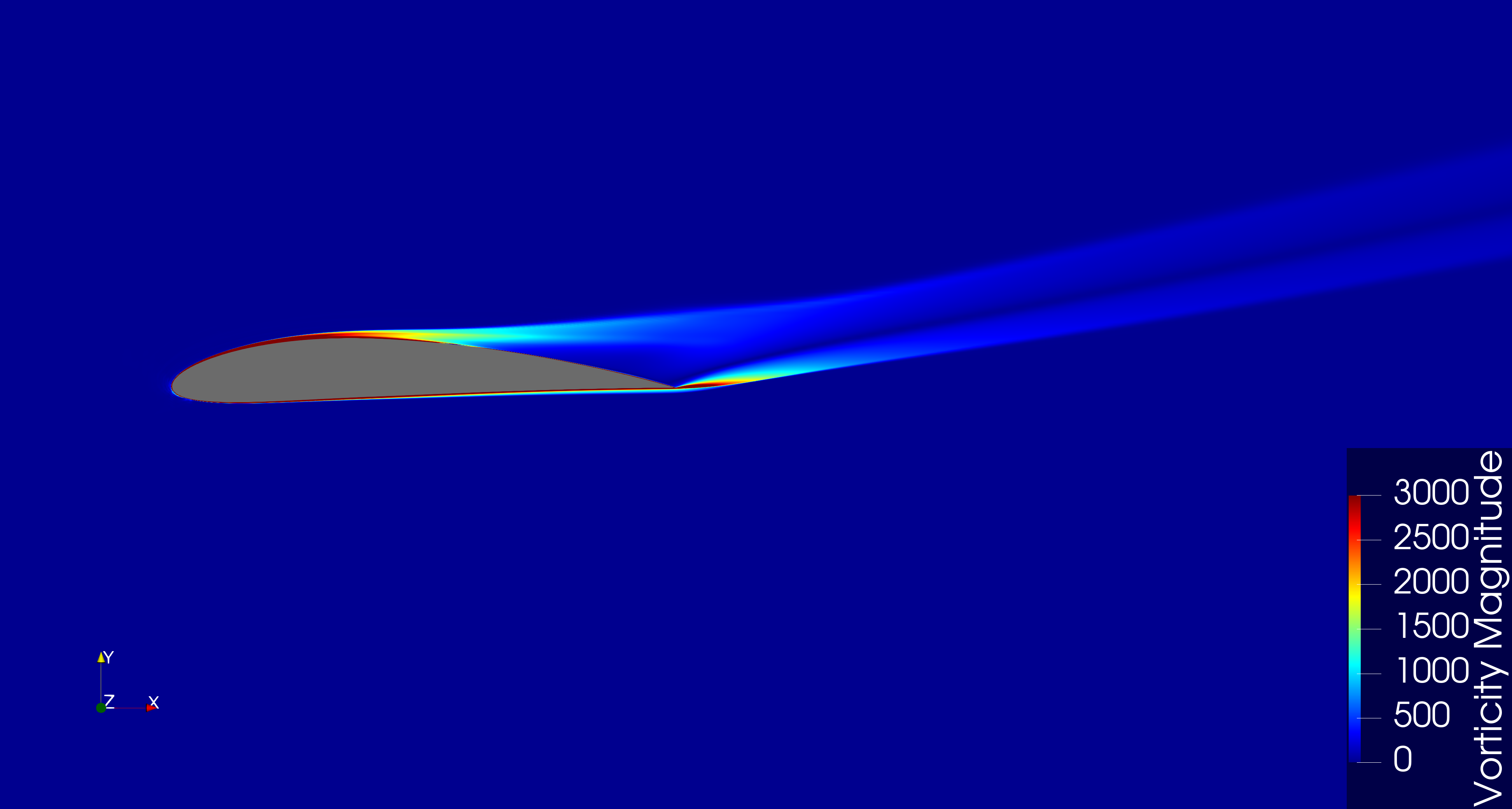}
        \caption{}
        \label{fig:WMCBF}
    \end{subfigure}
    \hfill
    \begin{subfigure}[b]{0.48\columnwidth}
        \centering
        \includegraphics[width=\linewidth]{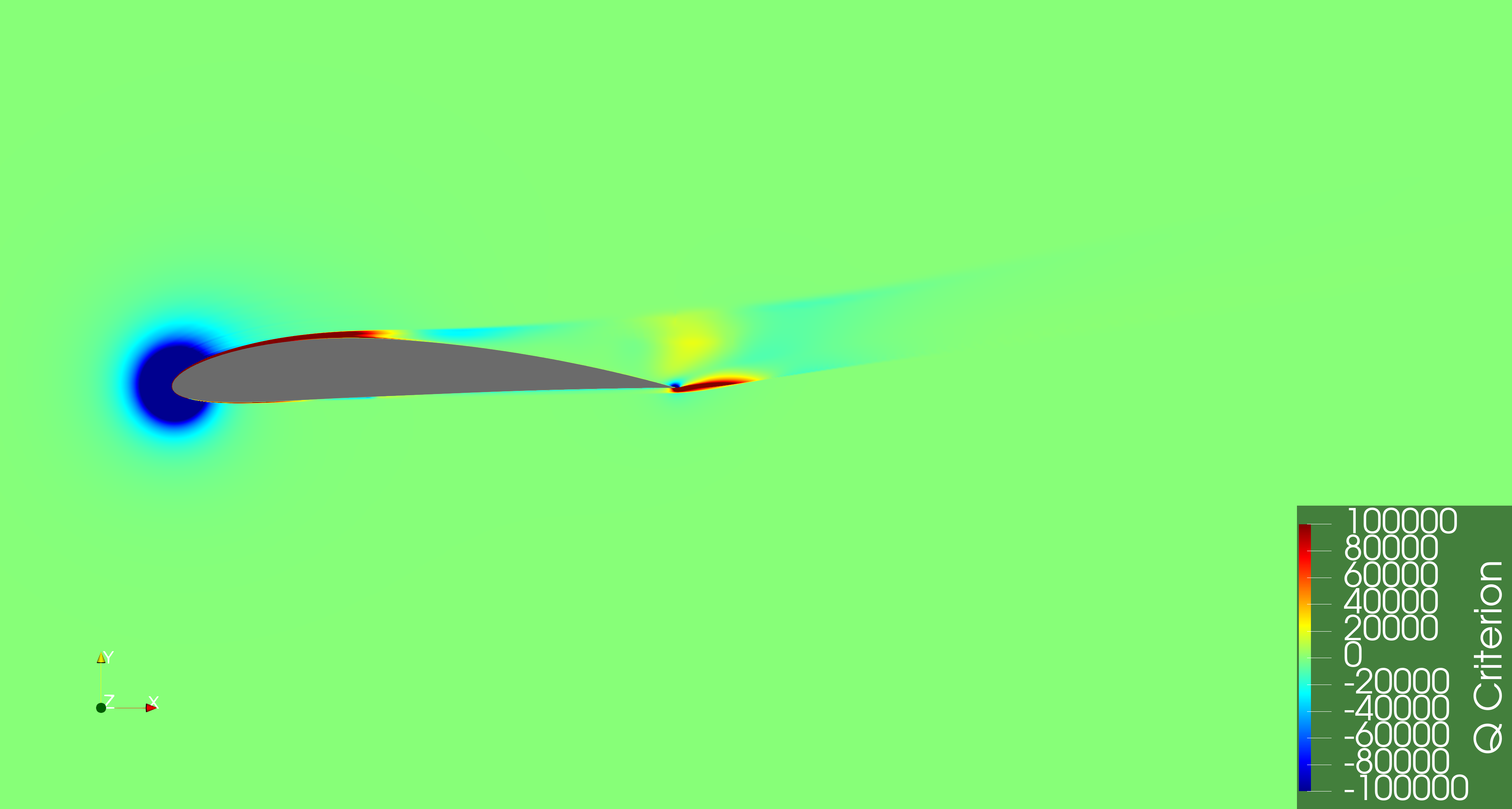}
        \caption{}
        \label{fig:QcritCBF}
    \end{subfigure}
    
    \caption{Contours of baseline flow: (a) Velocity magnitude [m/s], (b) Turbulent kinetic energy [m$^2$/s$^2$], (c) Vorticity magnitude [1/s], \textcolor{black}{and (d) Q-criterion [1/s$^2$]}. \textcolor{black}{Source: Authors' own work.}}
    \label{fig:CombinedBaselineFlowRow}
\end{figure}

\textcolor{black}{Fig.~\ref{fig:CombinedBaselineFlowRow} presents contours of different flow quantities in the baseline airfoil wake region, including velocity magnitude, turbulent kinetic energy (TKE), vorticity magnitude, and the Q-criterion. Based on the pressure coefficient distribution (Fig.~\ref{fig:Cpvsx}) and the velocity magnitude plot (Fig.~\ref{fig:VMCBF}), flow separation on the suction side occurs at a streamwise location of approximately \( 0.55c \).} Estimating the location of the flow separation point is essential for determining the appropriate placement of the actuator on the suction side. According to the study by Ebrahimi and Hajipour~\cite{B100}, placing the actuator upstream of the separation point is more effective in mitigating flow separation.

\textcolor{black}{The TKE distribution (Fig.~\ref{fig:TKECBF}) reveals the development of two free shear layers: one originating near the suction side separation point (approximately \( 0.55c \)), and the other from the trailing edge. These shear layers evolve into a turbulent wake, as indicated by elevated energy levels. This dual-shear-layer structure is also clearly observed in the vorticity magnitude and Q-criterion contours (Figs.~\ref{fig:WMCBF} and~\ref{fig:QcritCBF}, respectively). The presence of these two shear layers provides the physical basis for employing the dual-point excitation strategy, which targets simultaneous stimulation of both regions~\cite{B100}. Accordingly, in this study, the pressure side actuator is placed at the airfoil’s trailing edge.}

This baseline flow physics is a reference point for determining and evaluating the performance of various control strategies. Under baseline conditions, the lift coefficient (\( C_l \)) and drag coefficient (\( C_d \)) have been calculated as \textbf{\( C_l = 1.44 \)} and \textbf{\( C_d = 0.045 \)}, respectively. These values provide a quantitative measure of the aerodynamic performance of the airfoil in the absence of active control strategies.

\subsection{Adaptive MPC}

\textcolor{black}{The adaptive MPC was evaluated at two target lift coefficient values, \( C_l = 1.6 \) and \( 1.62 \), to assess its effectiveness in controlling flow separation. These target values (set-points) were selected based on prior open-loop flow control studies, as well as on the present analysis of the system's performance characteristics.} First, a target \(C_l\) of 1.6 was selected as the desired output. The objective was to adjust the excitation frequency of the plasma actuators in real time to achieve and maintain this lift coefficient. The control system dynamically responds to the airfoil's transient aerodynamic characteristics to ensure that the \(C_l\) reaches and stabilizes around the desired value of 1.6.

\begin{figure}[h!]
    \centering
    \includegraphics[width=0.9\columnwidth]{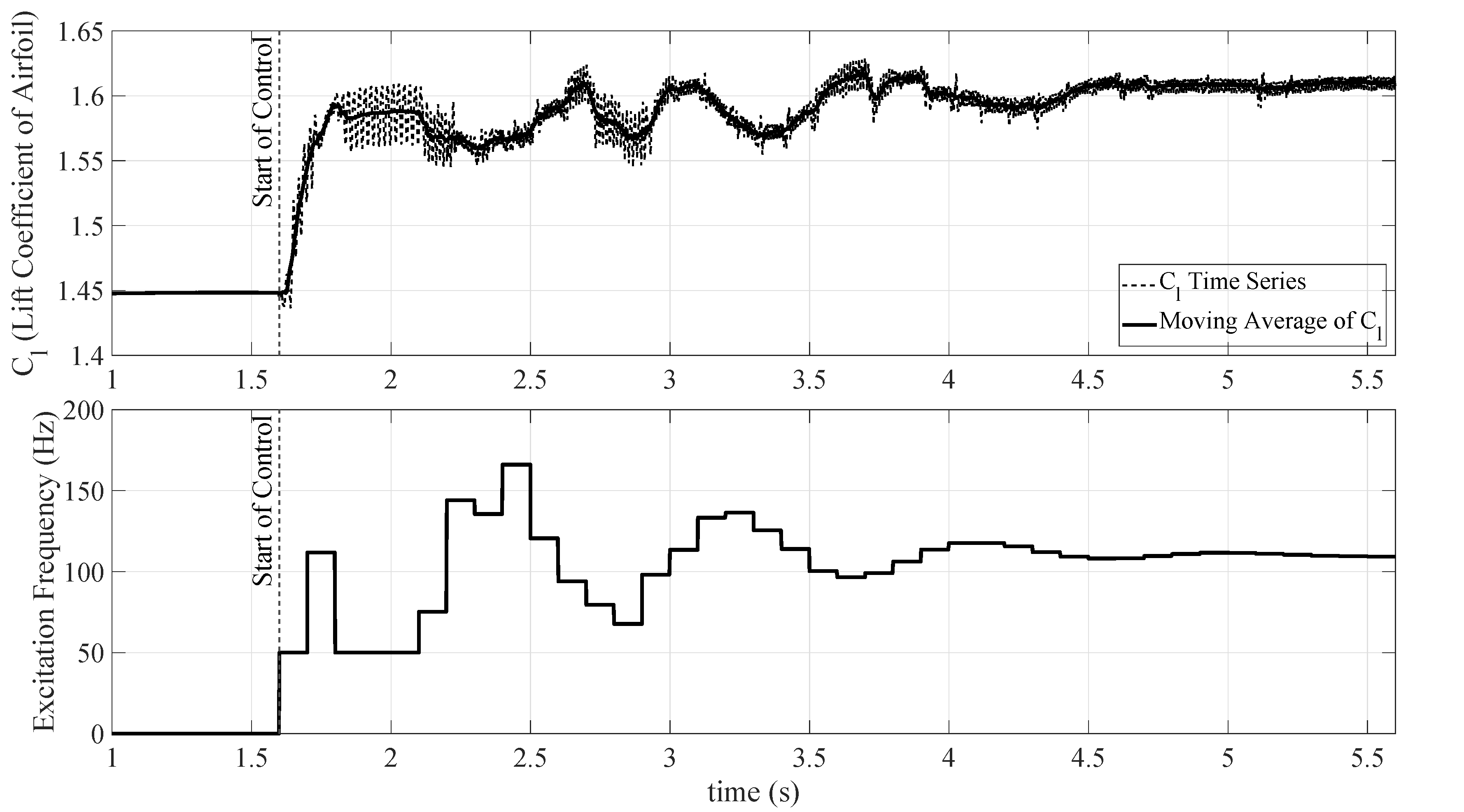}
    \caption{Desired \(C_l\) of 1.6 using adaptive MPC and its corresponding excitation frequencies. \textcolor{black}{Source: Authors' own work.}}
    \label{fig:eps_2}
\end{figure}

The graph presented in Fig.~\ref{fig:eps_2} illustrates the performance of the adaptive MPC system in achieving the desired lift coefficient. The top panel displays the time series of \( C_l \) values along with its moving average, while the bottom panel shows the corresponding excitation frequencies applied during the control process. The lift coefficient stabilizes approximately 2.8 seconds after the controller is activated. \textcolor{black}{According to the bottom panel, the actuator frequency converges to an excitation frequency of around 110~Hz. In terms of the normalized frequency, defined as \( F^+ = fc / U_{\infty} \), this corresponds to \( F^+ \approx 3 \).}

\begin{figure}[h!]
    \centering
    \includegraphics[width=0.9\columnwidth]{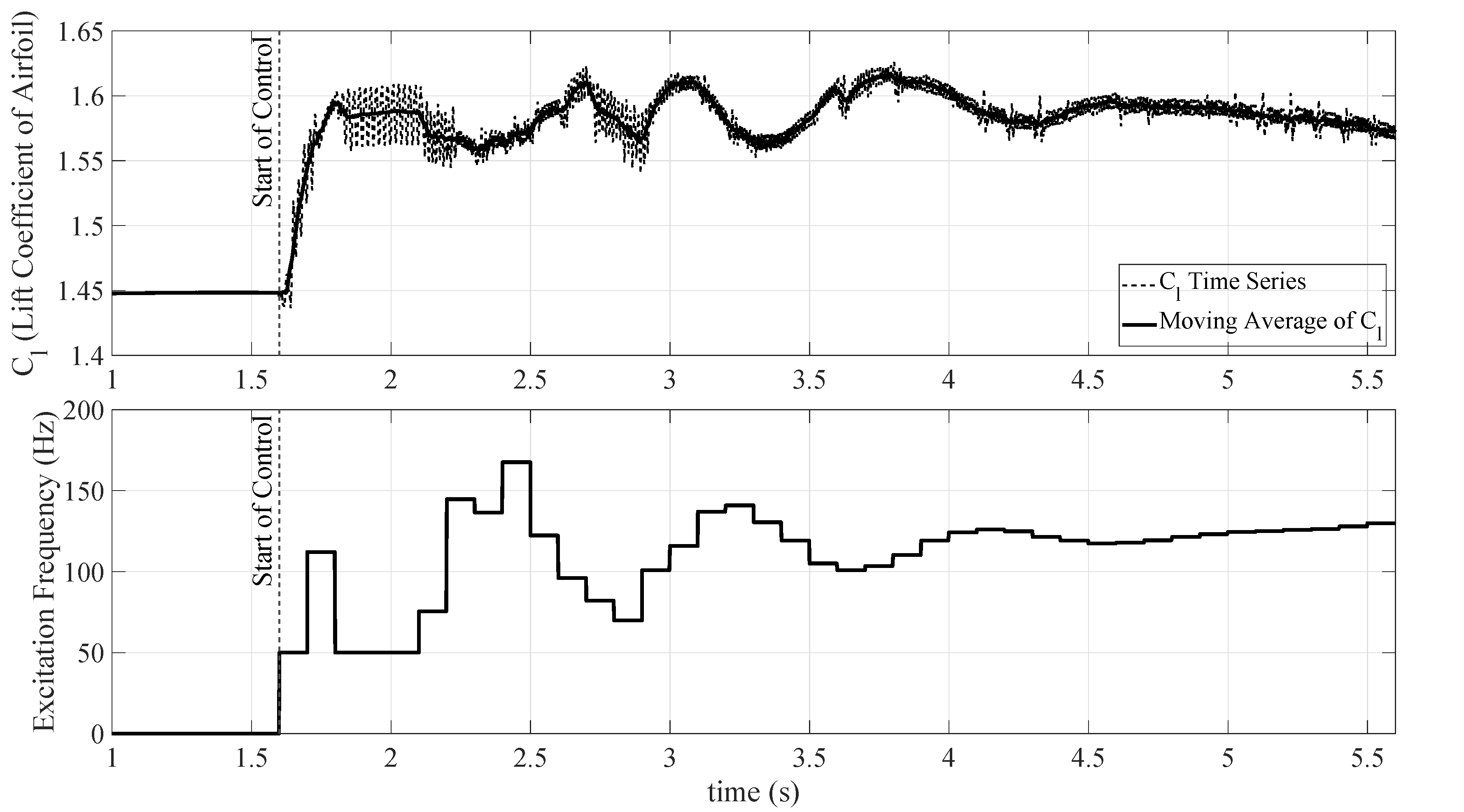}
    \caption{Desired \(C_l\) of 1.62 using Adaptive MPC and its corresponding excitation frequencies. \textcolor{black}{Source: Authors' own work.}}
    \label{fig:eps_3}
\end{figure}

To push the boundaries of the control system, aiming to achieve higher lift coefficients, a second target \(C_l\) of 1.62 was selected.  However, reaching such a high \(C_l\) presents significant challenges for the adaptive MPC to stabilize \(C_l\) around 1.62. As depicted in Fig.~\ref{fig:eps_3}, the controller could not stabilize the \(C_l\) at the desired value of 1.62. \textcolor{black}{The larger fluctuations in the \(C_l\) time series, compared with the previous case, reveal an unstable response as the controller struggles to regulate the set-point.}

\textcolor{black}{Unlike RL–based methods, which aim to maximize \(C_l\) through reward-driven optimization, the MPC framework is designed to track a predefined set-point. Consequently, MPC continuously attempts to reach and hold the specified \(C_l\). For the second target, \(C_l = 1.62\), this value lies near the physical limit of the dual plasma actuator configuration, making precise control difficult. Around \(t = 4.5\)~s, MPC increases the excitation frequency in an effort to attain the target, but this adjustment shifts the system away from the optimal frequency and results in a reduction in lift. Since the desired \(C_l\) exceeds what the actuators can sustainably produce, the controller fails to maintain the target, and the lift coefficient continues to oscillate below the set-point.}

\subsection{TDRL}

In contrast to the previous adaptive MPC approach, TDRL seeks to identify the optimal excitation frequency of plasma actuators that maximizes the \(C_l\). Here, the action space is discrete, meaning the control system can only select from a predefined set of excitation frequencies. The goal is to explore these discrete actions to find the frequency that yields the highest \(C_l\).

\begin{figure}[h!]
    \centering
    \includegraphics[width=0.9\columnwidth]{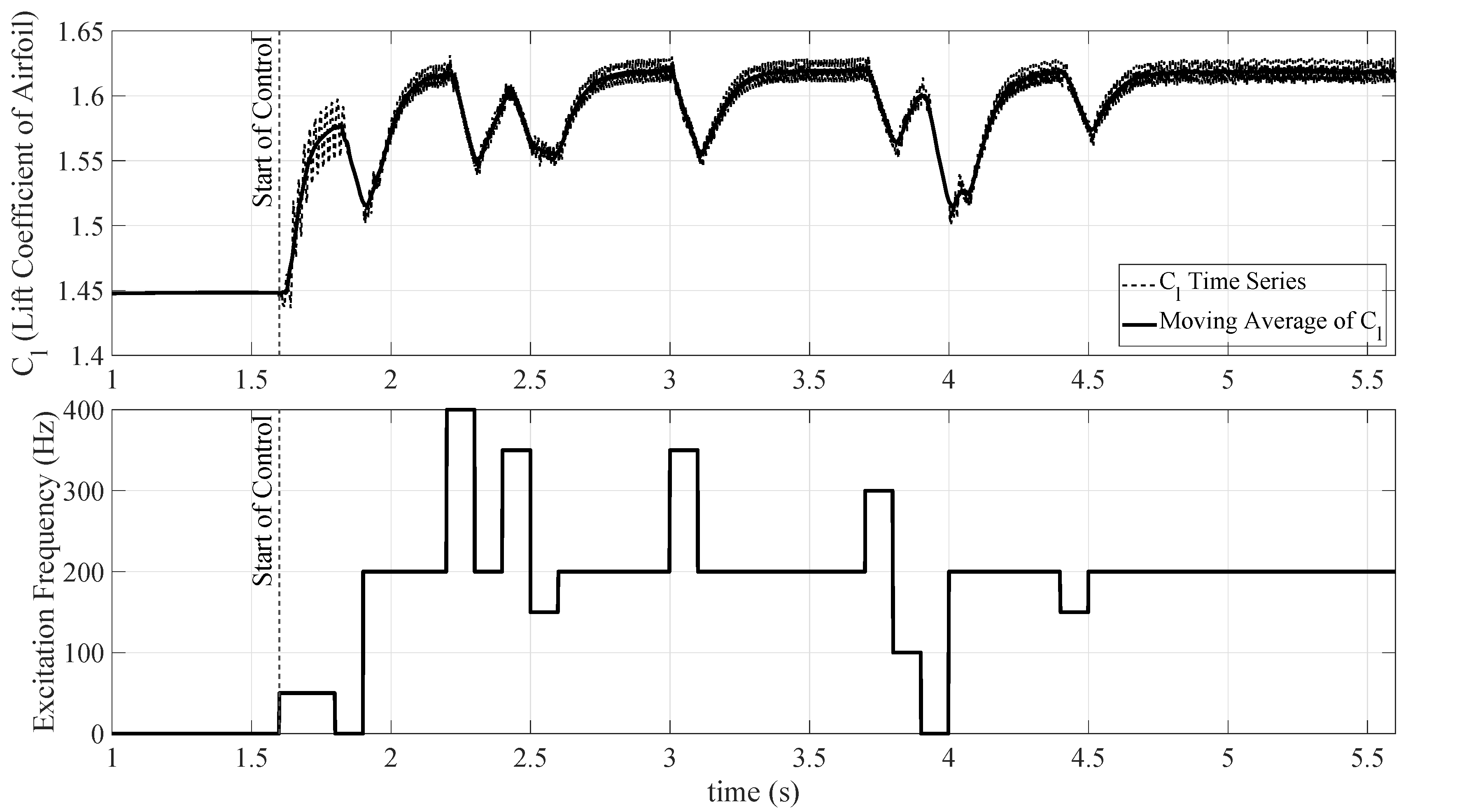}
    \caption{Exploring the optimal excitation frequency that maximizes \(C_l\) using TDRL. \textcolor{black}{Source: Authors' own work.}}
    \label{fig:eps_4}
\end{figure}

Based on Fig.~\ref{fig:eps_4}, the control system explored various frequencies before settling on 200 Hz as the optimal frequency for maximizing \(C_l\). The top panel shows how \(C_l\) evolves over time, while the bottom panel depicts the selected excitation frequencies throughout the process. By the end of the simulation, the system determined that 200 Hz was the most effective frequency for achieving the maximum \(C_l\). Notably, after four decision-makings, the controller found the optimal frequency of 200 Hz, resulting in a mean \(C_l\) of 1.619.

\subsection{DQL}

The DQL approach's control system is designed to learn the optimal policy for selecting excitation frequencies that maximize the \(C_l\). Unlike the previous TDRL, DQL utilizes a neural network to estimate the Q-values for each action, enabling the system to make more informed decisions in complex environments. The action space remains discrete, and the system learns by exploring different excitation frequencies and updating its policy based on the observed outcomes.

\begin{figure}[h!]
    \centering
    \includegraphics[width=0.9\columnwidth]{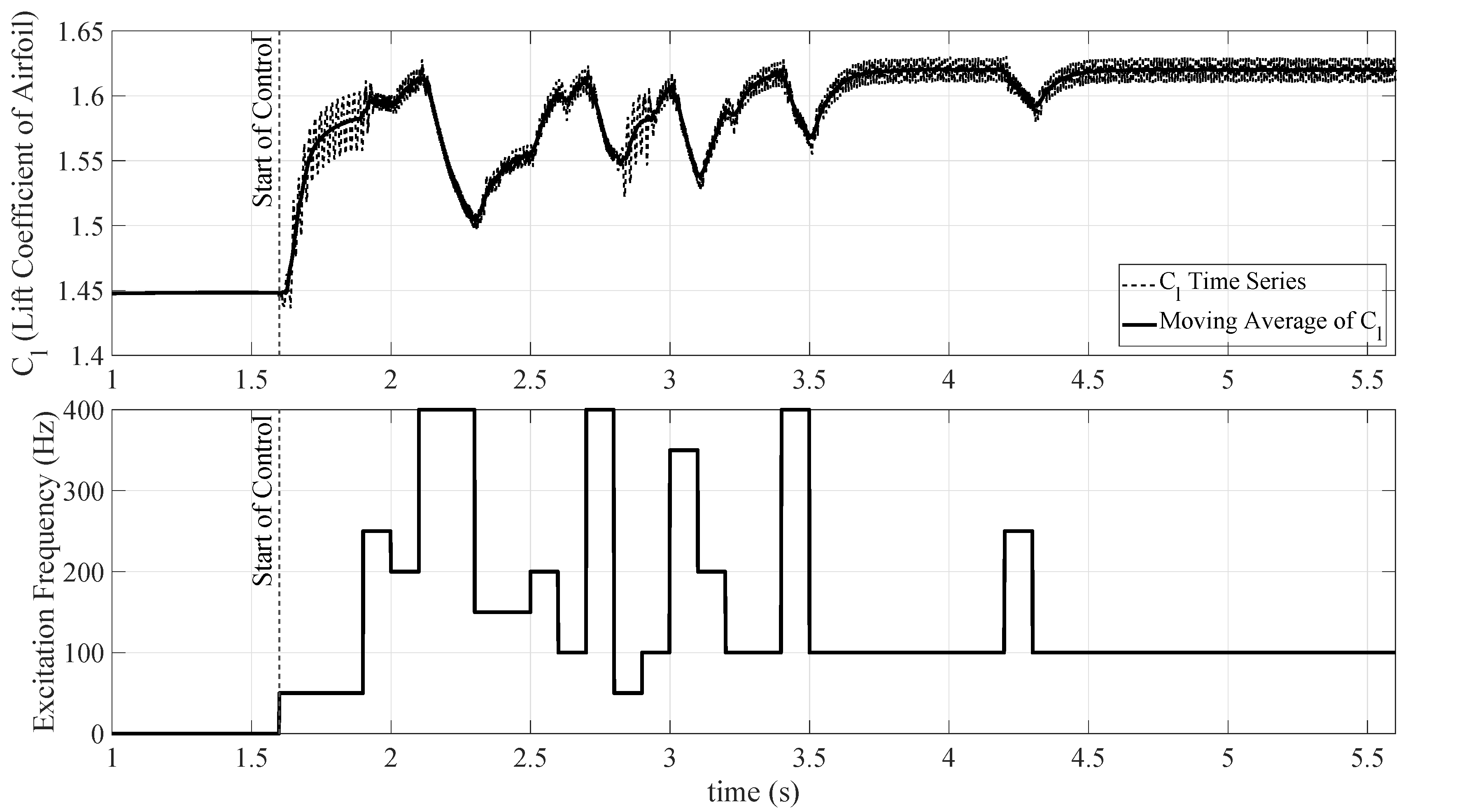}
    \caption{Selection of the optimal excitation frequency for maximizing \(C_l\) using DQL. \textcolor{black}{Source: Authors' own work.}}
    \label{fig:eps_5}
\end{figure}

As shown in Fig.~\ref{fig:eps_5}, the DQL algorithm explored various frequencies before selecting 100 Hz as the optimal excitation frequency. Despite the fluctuations during the learning phase, the system eventually converged to 100 Hz as the most effective frequency for achieving a high and stable \(C_l\). After 11 desicion makings, the controller found the optimal frequency of 100 Hz, resulting in a mean \(C_l\) of 1.619.

\subsection{Signal processing integrated with DQL}

\begin{figure}[h!]
    \centering
    \includegraphics[width=0.9\columnwidth]{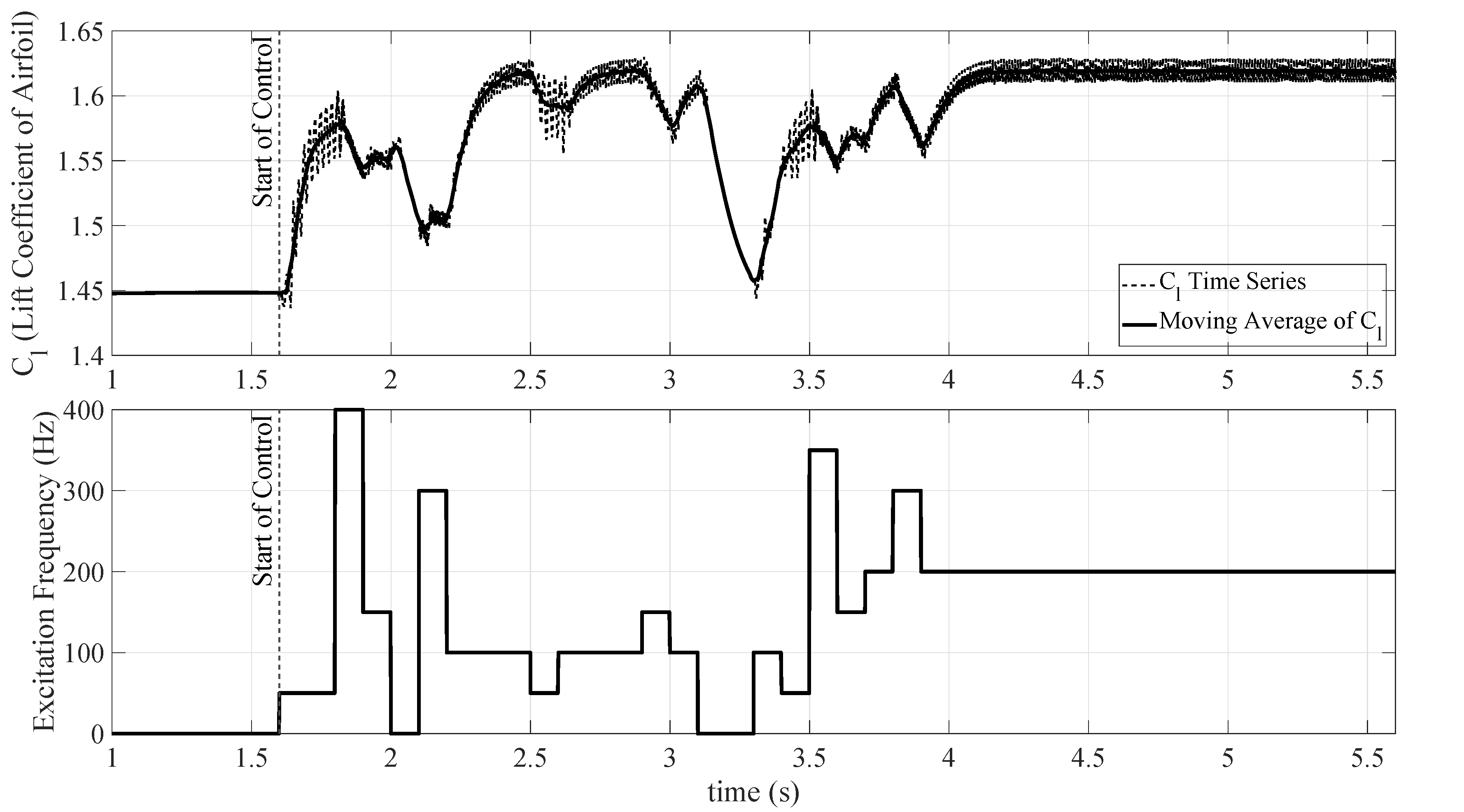}
    \caption{Selection of the optimal excitation frequency for maximizing \(C_l\) using signal processing integrated with DQL. \textcolor{black}{Source: Authors' own work.}}
    \label{fig:eps_6}
\end{figure}

In this approach, DQL was enhanced by integrating signal processing techniques to improve the accuracy and robustness of the excitation frequency selection process. The objective was to refine the control system's ability to identify the optimal frequency for maximizing the lift coefficient. According to Fig.~\ref{fig:eps_6}, the system eventually selected 200 Hz as the optimal frequency. After 21 desicion makings, the controller found the optimal frequency of 200 Hz, resulting in a mean \(C_l\) of 1.619.

\begin{figure}[h!]
    \centering
    \includegraphics[width=0.9\columnwidth]{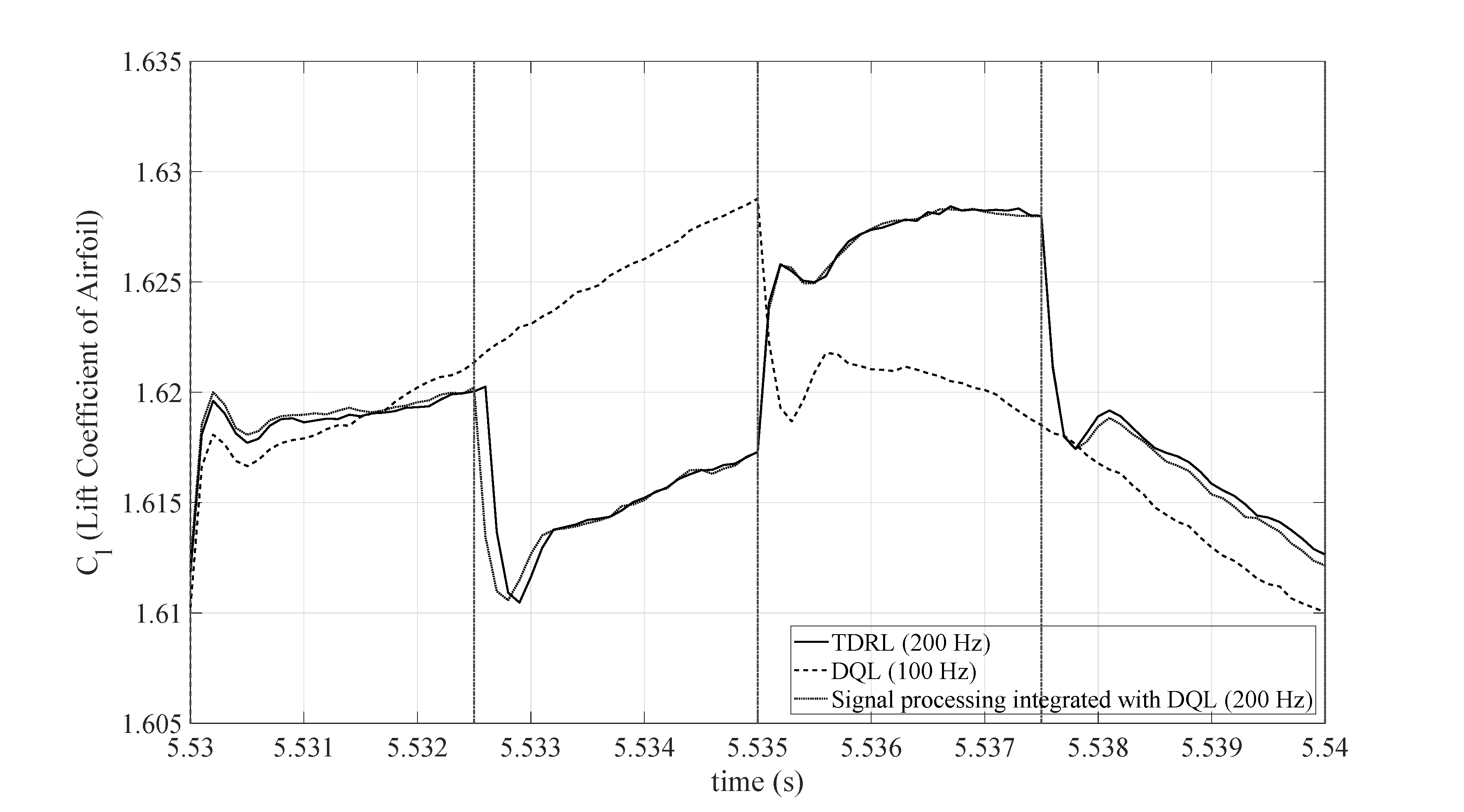}
    \caption{Comparison of results from different reinforcement learning methods for determining \(C_l\). \textcolor{black}{Source: Authors' own work.}}
    \label{fig:ComparisonRL}
\end{figure}

The average \(C_l\) for 100 Hz and 200 Hz are nearly identical, which made the decision-making process challenging for the network. The reason they are similar can be seen in the Fig.~\ref{fig:ComparisonRL}. For 100 Hz, there is one period of ``on" and ``off" states, while for 200 Hz, there are two periods. However, the first ``on" phase for 200 Hz has a lower maximum than the ``on" phase for 100 Hz, and the second ``on" phase for 200 Hz reaches a higher maximum than 100 Hz. As a result, the average of two consecutive ``on" states at 200 Hz closely matches the average of a single ``on" state at 100 Hz.

This behavior suggests that the DQL produces similar average \(C_l\) compared to the signal processing integrated with DQL and TDRL cases, making it harder for the network to differentiate between the two based purely on \(C_l\) values. \textcolor{black}{This similarity in average \(C_l\) values helps explain why TDRL and signal processing integrated with DQL methods converged to 200 Hz, as their initial exploration behavior within the discrete action space, guided by early random sampling and the reward landscape, may have favored 200 Hz slightly, whereas DQL’s function approximator ultimately selected 100 Hz.}

\subsection{Comparative Analysis}

In this section, we compare different control strategies applied to the problem of flow separation control over a static airfoil at a near-stall angle of attack condition. Based on the algorithms discussed, the complexity level of the RL-based methods increases in the following order: TDRL, DQL, and signal processing integrated with DQL. As the complexity of RL-based methods increases, they require more decision-making steps for accurate system identification.

\begin{figure}[h!]
    \centering
    \includegraphics[width=0.9\columnwidth]{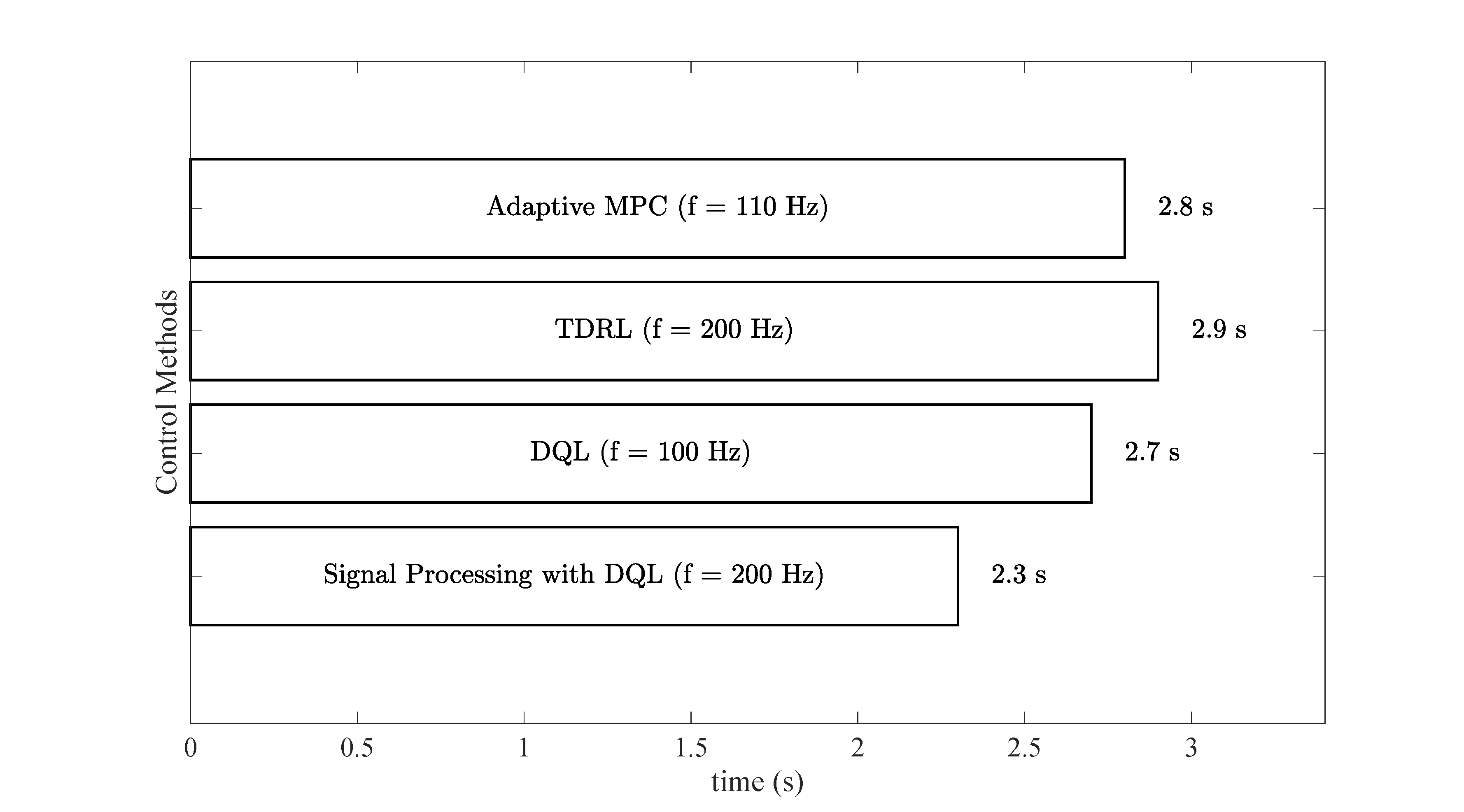}
    \caption{Stabilization time and converged frequencies of different control methods. \textcolor{black}{Source: Authors' own work.}}
    \label{fig:Comparison}
\end{figure}

As demonstrated in Fig.~\ref{fig:Comparison}, the time needed to achieve stability is reduced despite the higher number of decisions required by more complex methods. Also, it is observed that all RL-based methods successfully converge to an optimal \(C_l\), albeit with different choices of actuation frequencies, achieving similar final values. In the case of the adaptive MPC approach, targeting a lift coefficient of 1.6, the actuator frequency stabilized around the excitation frequency of 110 Hz, which corresponds to \(F^+ \approx 3\). The chosen target lift coefficient in the MPC method is close to the optimal value identified by RL-based methods, leading to an excitation frequency similar to that achieved by the DQL method (i.e., 100 Hz). This indicates that despite the differences in approach, there is a convergence in the key parameters, particularly in excitation frequency, between the adaptive MPC and RL-based methods.

\textcolor{black}{\subsection{Aerodynamic Characteristics of the Controlled Flow}}

\begin{figure}[h!]
    \centering
    \begin{subfigure}[b]{0.48\columnwidth}
        \centering
        \includegraphics[width=\linewidth]{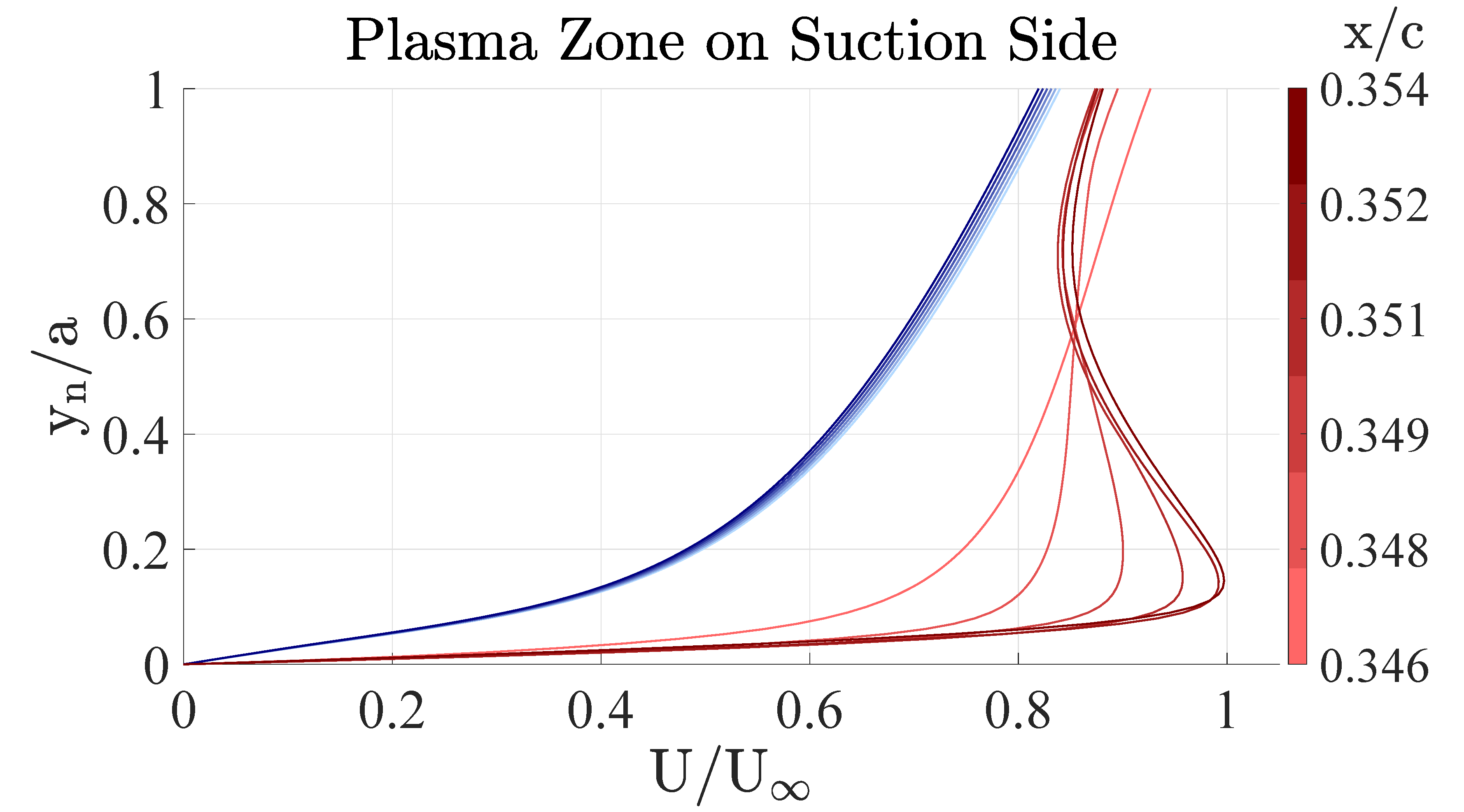}
        \caption{}
        \label{fig:SSplasma}
    \end{subfigure}
    \hfill
    \begin{subfigure}[b]{0.48\columnwidth}
        \centering
        \includegraphics[width=\linewidth]{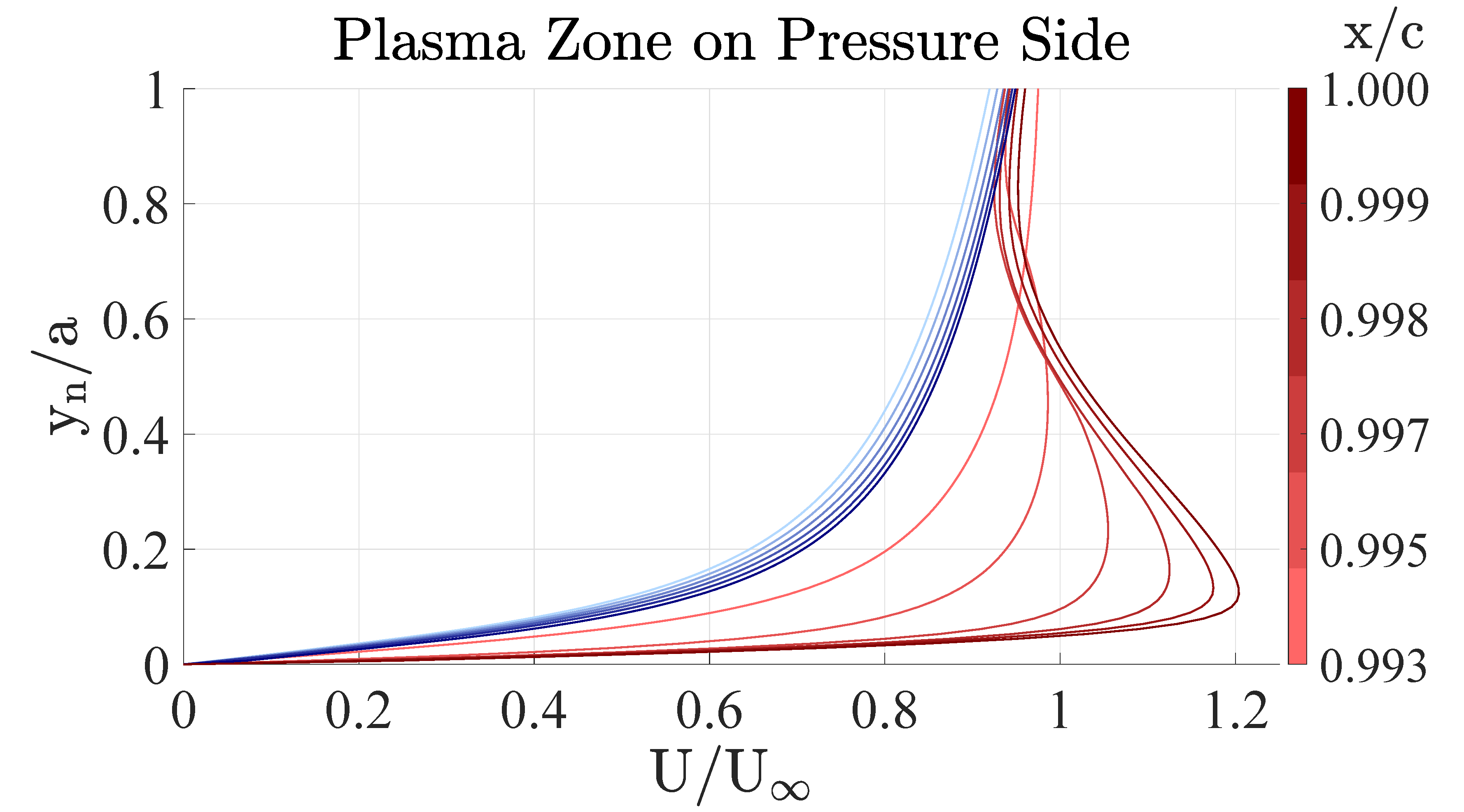}
        \caption{}
        \label{fig:PSplasma}
    \end{subfigure}

    \caption{\textcolor{black}{Velocity profiles within the plasma actuator zones on the suction and pressure sides, shown with and without plasma actuation. Red and blue shading represent plasma-on and plasma-off conditions, respectively; darker shades indicate downstream streamwise locations.} \textcolor{black}{Source: Authors' own work.}}
    
    \label{fig:plasmaOnOff}
\end{figure}

\textcolor{black}{
In this subsection, we examine the flow behavior under active control with dual-point excitation of plasma actuators. Fig.~\ref{fig:plasmaOnOff} illustrates the effect of plasma actuation on boundary layer velocity profiles within the plasma zones on both the suction and pressure sides of the airfoil. The horizontal axis represents the velocity magnitude normalized by the freestream velocity, while the vertical axis denotes the wall-normal position (\(y_n\)) normalized by the plasma zone height (\(a\)). Each plot includes twelve velocity profiles sampled at six streamwise locations within the plasma zones, depicted using a color gradient where darker shades correspond to greater \(x/c\) positions. Red-shaded profiles indicate the plasma-on case with the DQL-integrated signal processing method, while blue-shaded profiles represent the plasma-off condition at the same streamwise locations.}

\textcolor{black}{In the absence of plasma actuation, the boundary layer develops in a typical manner across the actuator regions. Notably, the velocity profiles on the suction side (Fig.~\ref{fig:SSplasma}) exhibit lower fullness compared to those on the pressure side (Fig.~\ref{fig:PSplasma}), due to the influence of an adverse pressure gradient on the suction side versus a favorable pressure gradient on the pressure side. Upon activation of the plasma actuators, an increase in profile fullness is observed at the upstream positions within the plasma zones, indicating momentum enhancement due to the added body force. Further downstream, the development of an ionic wind alters the boundary layer into a wall-jet-like profile with significantly higher momentum relative to the inactive case. Since the actuators operate in a periodic excitation mode, this results in the periodic formation and decay of the ionic wind. This mechanism plays a central role in generating instabilities in the flow field, which is a key feature of the active control strategy.}

\begin{figure}[h!]
    \centering
    \begin{subfigure}[b]{0.48\columnwidth}
        \centering
        \includegraphics[width=\linewidth]{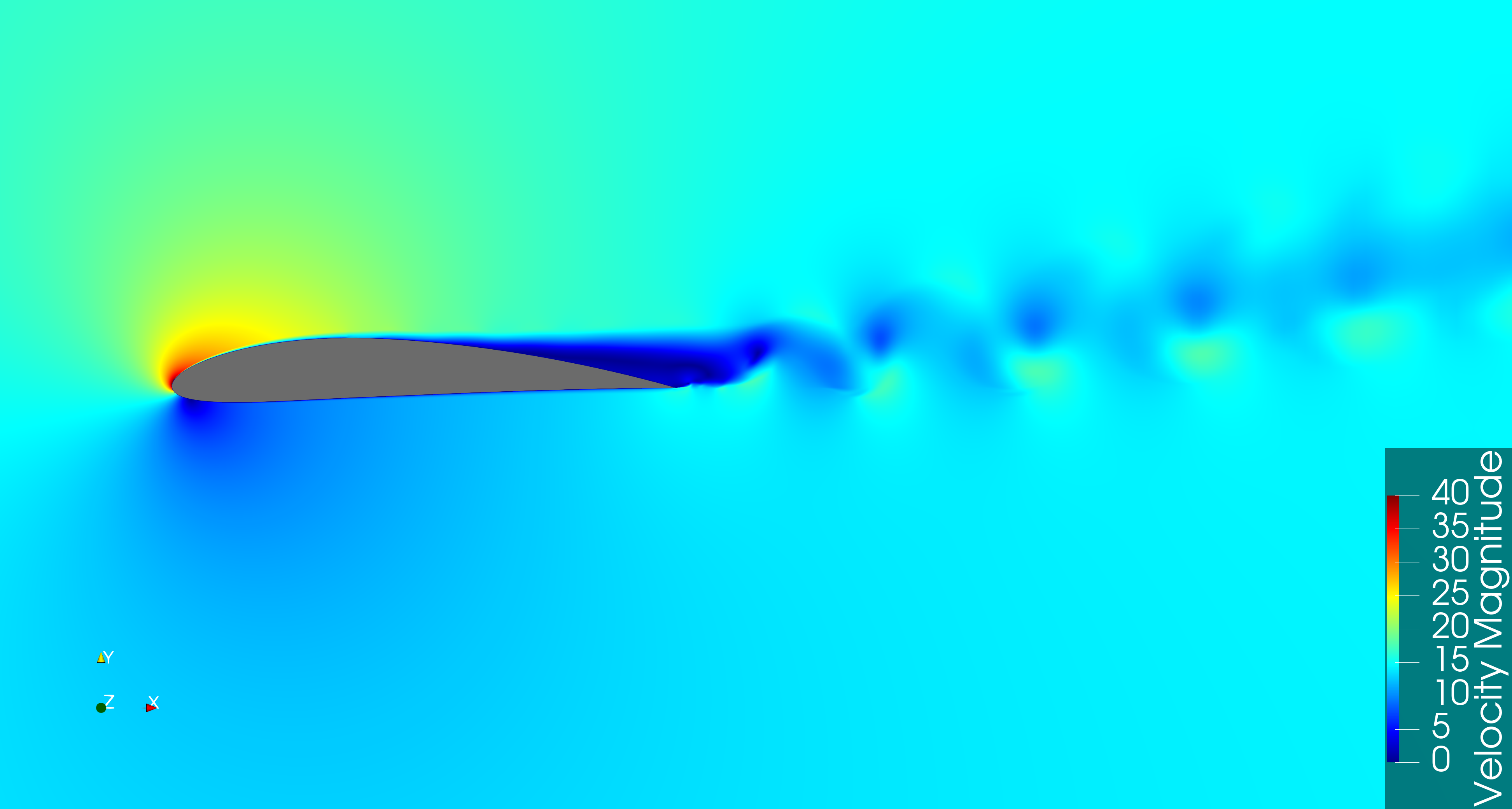}
        \caption{}
        \label{fig:VMCBF2}
    \end{subfigure}
    \hfill
    \begin{subfigure}[b]{0.48\columnwidth}
        \centering
        \includegraphics[width=\linewidth]{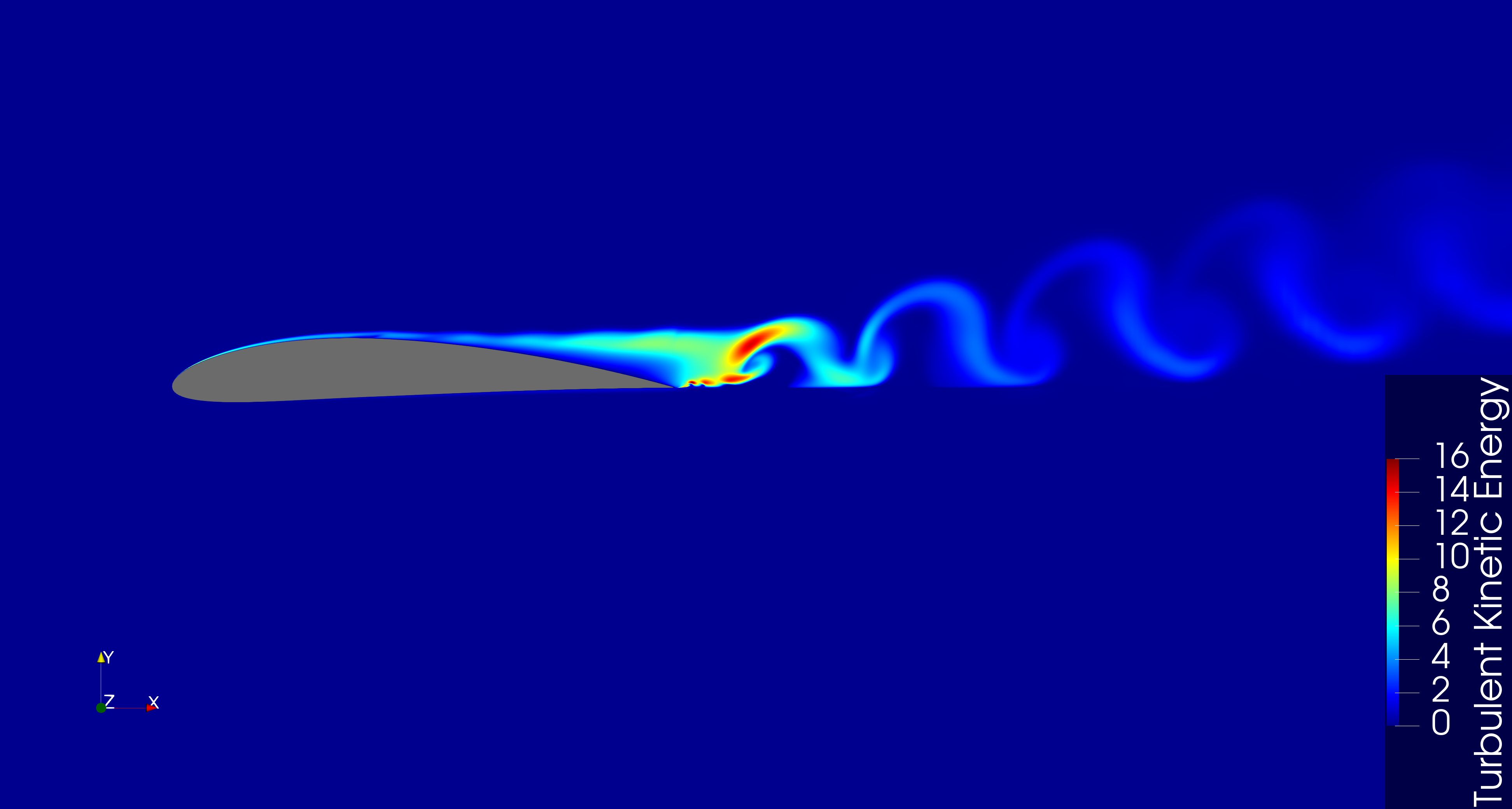}
        \caption{}
        \label{fig:TKECBF2}
    \end{subfigure}
    
    \begin{subfigure}[b]{0.48\columnwidth}
        \centering
        \includegraphics[width=\linewidth]{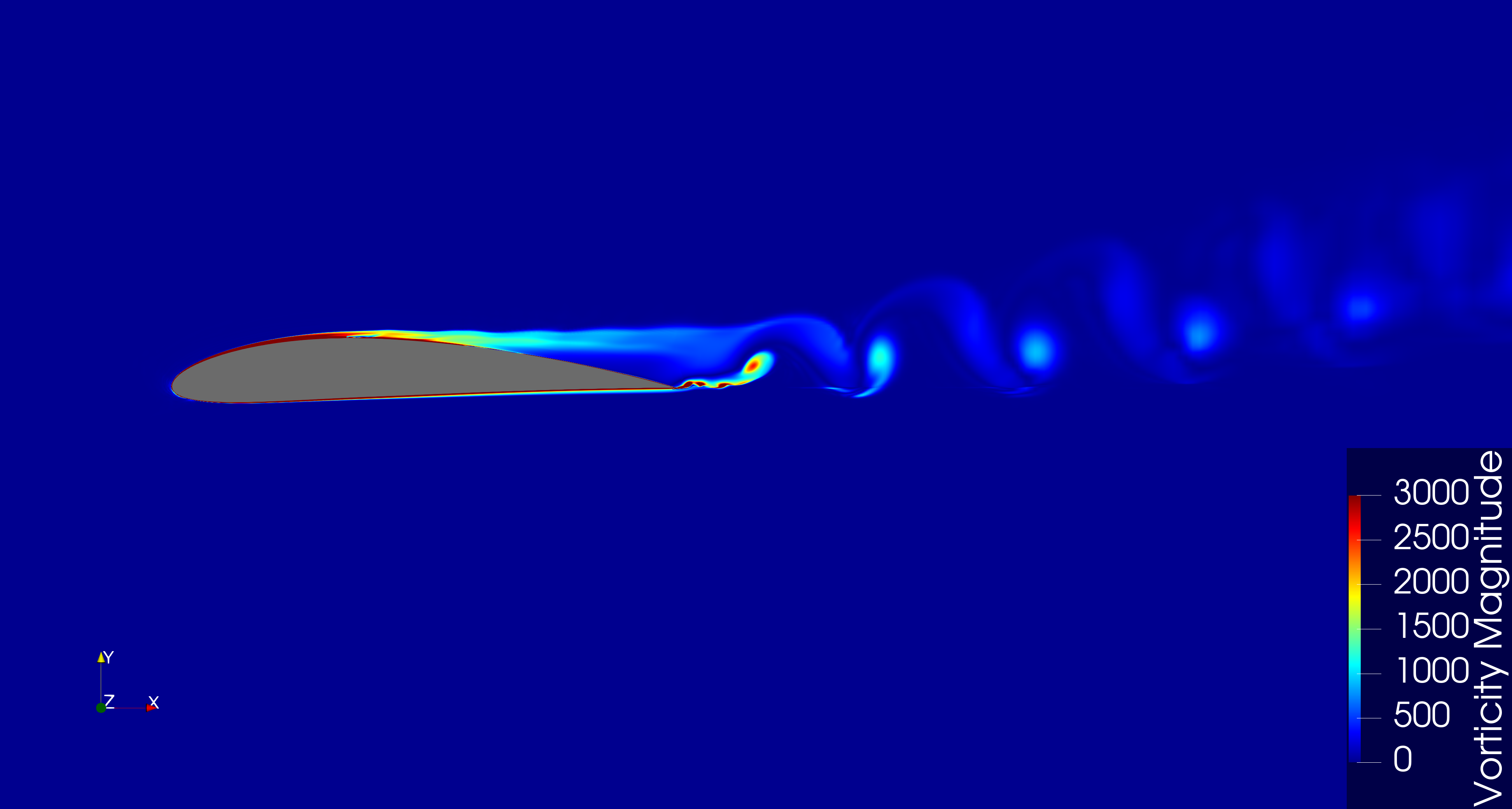}
        \caption{}
        \label{fig:WMCBF2}
    \end{subfigure}
    \hfill
    \begin{subfigure}[b]{0.48\columnwidth}
        \centering
        \includegraphics[width=\linewidth]{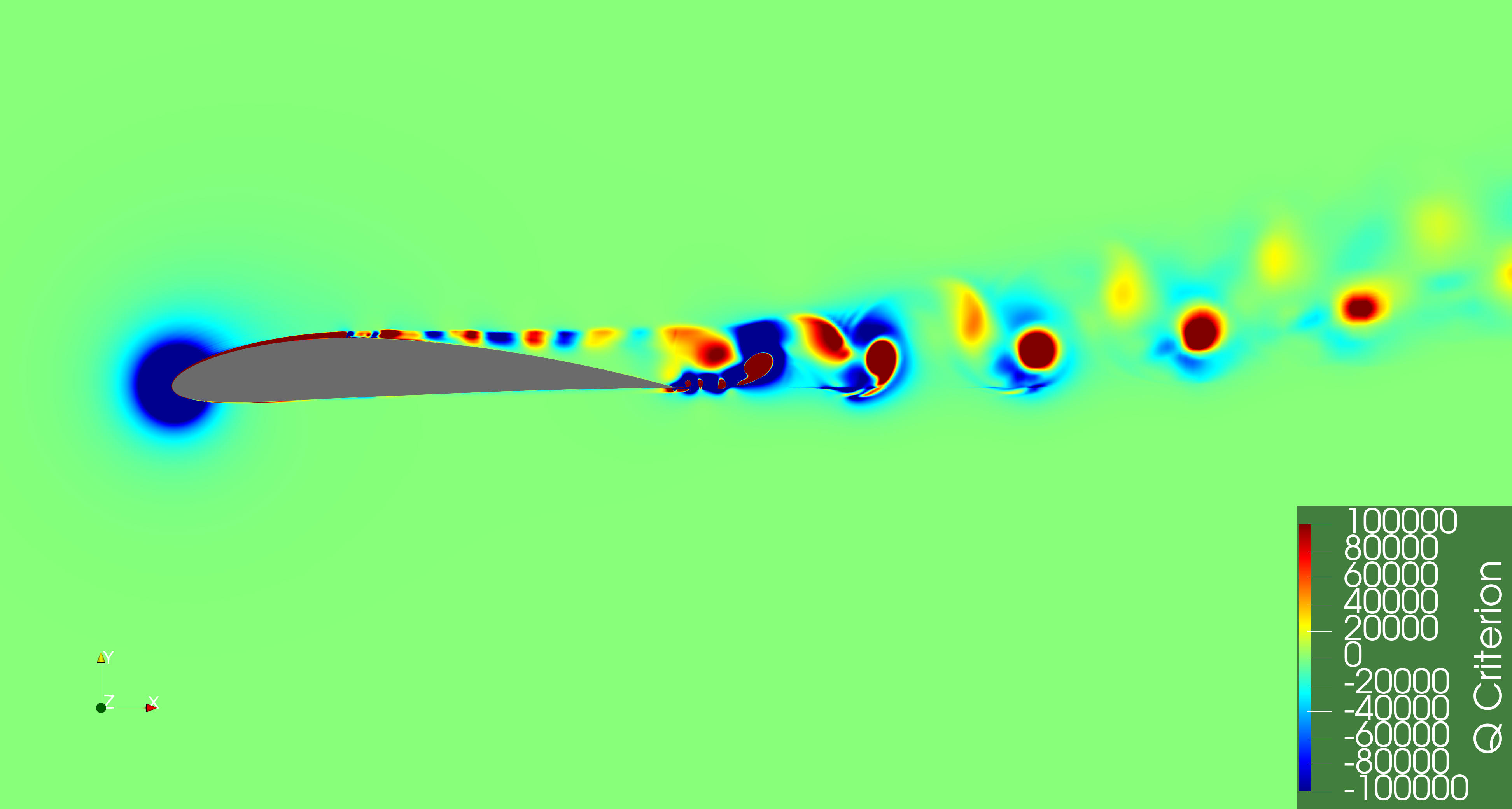}
        \caption{}
        \label{fig:QcritCBF2}
    \end{subfigure}
    
    \caption{Contours of controlled flow using signal processing integrated with DQL method: (a) Velocity magnitude [m/s], (b) Turbulent kinetic energy [m$^2$/s$^2$], (c) Vorticity magnitude [1/s], and \textcolor{black}{(d) Q-criterion [1/s$^2$]}. \textcolor{black}{Source: Authors' own work.}}
    \label{fig:CombinedControlledFlowGrid}
\end{figure}

To compare the controlled flow field with the baseline flow (as shown in Fig.~\ref{fig:CombinedBaselineFlowRow}), Fig.~\ref{fig:CombinedControlledFlowGrid} presents the contours of velocity magnitude, turbulent kinetic energy, vorticity magnitude, \textcolor{black}{Q-criterion}. These results are specifically provided for the signal processing integrated with DQL method. Since the flow fields for other control methods are nearly similar, their contours are not presented. This figure demonstrates that the proposed algorithms have successfully modified the flow field, including shear layers, wake, and separation points, compared to the baseline case. 

The velocity magnitude contours in Figs. \ref{fig:VMCBF} and \ref{fig:VMCBF2} reveal a reduction in the extent of the separation zone under the influence of the flow control method. \textcolor{black}{This observation is further supported by the reduced TKE levels in the suction-side separated shear layer of the controlled case compared to the baseline flow (Figs.~\ref{fig:TKECBF} and \ref{fig:TKECBF2}). Downstream of the trailing edge, the shear layer turns into discrete turbulent vortices.} Additionally, the vorticity magnitude contours in Fig.~\ref{fig:WMCBF} and \ref{fig:WMCBF2} indicate that the separated shear layer on the airfoil’s suction side is inclined closer to the airfoil surface compared to the baseline case. \textcolor{black}{This behavior promotes greater entrainment of high-energy free-stream fluid into the separated region on the suction side, thereby contributing to a reduction in the size of the separation zone. Furthermore, the coherent vorticity structures observed in the wake of the controlled flow suggest the presence of a periodic vortex-shedding pattern, attributed to the excitation of the shear layers.}

\textcolor{black}{The Q-criterion contours in Fig.~\ref{fig:QcritCBF2} provide deeper insight into the influence of plasma actuator excitation on the development of the separated shear layer, compared to the baseline configuration shown in Fig.~\ref{fig:QcritCBF}. Just downstream of the DBD plasma actuator (about the suction-side separation point) the controlled case exhibits alternating coherent regions of positive and negative Q values, indicating zones of elevated vorticity and strain rate, respectively. This pattern signifies the dominance of instabilities within the shear layer under plasma actuation. Moreover, the observed wavelength of these instabilities is synchronized with the excitation frequency of the actuator, a phenomenon known as lock-on, wherein shear layer dynamics resonate with external periodic forcing~\cite{samimy2018excitation}. These synchronized instabilities accelerate the transition of the shear layer into turbulent, discrete vortices, thereby enhancing mixing between the separated region and the surrounding flow.}

\begin{figure}[h!]
    \centering
    \includegraphics[width=0.8\columnwidth]{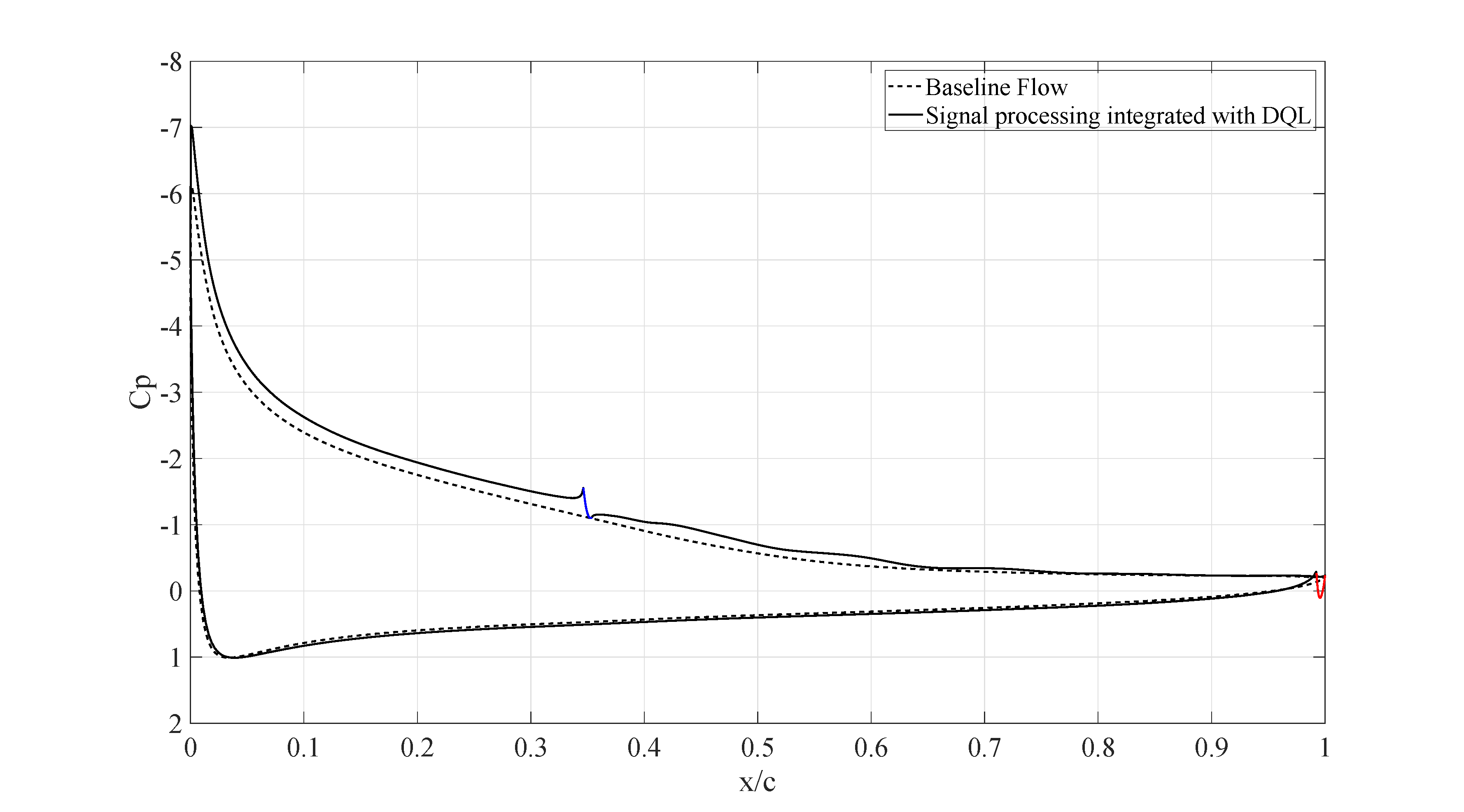}
    \caption{\textcolor{black}{Comparison of \(C_p\) distribution along \(x/c\) between baseline flow and signal processing integrated with DQL.} \textcolor{black}{Source: Authors' own work.}}
    \label{fig:CpvsxComp}
\end{figure}

\textcolor{black}{Fig.~\ref{fig:CpvsxComp} compares the pressure coefficient (\(C_p\)) distribution along the airfoil surface for the baseline and controlled case. The locations of the two plasma actuators are indicated in red and blue, corresponding to the pressure and suction sides of the airfoil, respectively. On the pressure side, the control case exhibits slightly higher pressure values compared to the baseline. However, the overall difference in this region is not significant, as the flow remains largely attached. In contrast, the suction side shows a marked difference due to the applied flow control. From \(x/c = 0\) to approximately \(x/c = 0.65\), the controlled case exhibits a lower \(C_p\) (i.e., higher suction) relative to the baseline. This increase in suction on the upper surface, coupled with the marginally elevated pressure on the lower surface, leads to a greater pressure differential across the airfoil and consequently results in an increased lift coefficient.}

\section{Conclusions}

In this study, we explored various advanced control strategies for optimizing the aerodynamic performance of an airfoil by managing flow separation using plasma actuators. The concept previously introduced by the authors, dual-point excitation, was employed to simultaneously excite both the suction side and trailing-edge shear layers using DBD plasma actuators. The applied control methods include adaptive MPC, TDRL, DQL, and integrated signal processing with the DQL method. These techniques were assessed based on their ability to stabilize the lift coefficient at targeted values under different operating conditions. Our findings revealed insights into the effectiveness and limitations of these methods in a highly dynamic and non-linear environment. The main findings are summarized as follows:

\begin{itemize}
    \item \textbf{Adaptive MPC Performance:}
    \begin{itemize}
        \item The adaptive MPC approach effectively achieved the desired lift coefficient, $C_l$, of 1.60, which is within the system's physical limitations.
        \item Instabilities were observed when attempting to push $C_l$ more than the upper physical limits, highlighting challenges in controlling the system near the actuator's maximum capacity.
        \item While adaptive MPC is effective for moderate targets, its performance may deteriorate when extreme precision and control are required in more demanding situations.
    \end{itemize}
    
    \item \textbf{RL-Based Methods:}
    \begin{itemize}
        \item RL-based methods, including TDRL, DQL, and DQL integrated with signal processing, demonstrated effective selection of excitation frequencies for maximizing $C_l$.
        \item Challenges were noted in distinguishing between the optimal value of two excitation frequencies (100 Hz and 200 Hz) due to near-identical mean $C_l$ values achieved at these frequencies.
    \end{itemize}
    
    \item \textbf{Convergence and Frequency Observations:}
    \begin{itemize}
        \item The time required to achieve \(C_l\) value convergence is reduced despite the higher number of decisions necessitated by more complex methods.
        \item All RL-based methods successfully converged to an optimal $C_l$, albeit with different excitation frequencies (Fig.~\ref{fig:Comparison}).
        \item The adaptive MPC approach stabilized at an actuator frequency of approximately 110 Hz ($F^+ \approx 3$) for a target $C_l$ of 1.6, aligning closely with the excitation frequency identified by the DQL method.
        \item There is a convergence in key parameters, such as excitation frequency, between adaptive MPC and RL-based methods, underscoring the alignment of both approaches in determining effective actuation strategies.
    \end{itemize}
\end{itemize}

\textcolor{black}{Further, this study offers new insights into active flow control through the implementation of dual-point DBD plasma actuation. The interaction between this actuation strategy and advanced control techniques provides an effective contribution to mitigating flow separation near stall conditions. The improved aerodynamic performance observed in the controlled cases is primarily attributed to the amplification of shear layer instabilities, which enhance flow entrainment and effectively delay separation. Consequently, a larger suction peak and a more favorable pressure gradient are established over the upper surface of the airfoil, leading to a higher lift coefficient.}

\textcolor{black}{From a physical perspective, the differences in control performance between MPC and RL can be linked to how each method handles the unsteady flow dynamics near stall, where nonlinear flow features dominate. The adaptive MPC controller maintains good performance up to a threshold ($C_L = 1.6$); however, beyond this point, it struggles to ensure stability as the plasma actuator approaches its physical limitations. In contrast, the RL controller adapts more effectively to variations in the separated shear layers, successfully suppressing flow separation and enabling sustained, stable lift enhancement under near-stall conditions.}

Future work could improve turbulence modeling and enhance the control system's ability to detect subtle frequency effects, potentially leading to more precise and effective flow control strategies. Additionally, incorporating the pitch and plunge motions of the airfoil could provide a basis for more complex aerodynamic problems to evaluate these control techniques.

\section*{Data Availability Statement}

The data that support the findings of this study are available from the corresponding author upon reasonable request.

\bibliographystyle{elsarticle-num} 
\bibliography{ref.bib}

\end{document}